\documentclass[review]{elsarticle}

\usepackage{lineno,hyperref}
\usepackage[a4paper, total={7in, 9in}]{geometry}
\usepackage{graphicx}
\usepackage{chemformula}
\usepackage{amsmath,empheq,amsthm,amssymb}
\usepackage[titletoc,title,page]{appendix}
\usepackage{mathtools}
\usepackage{xparse}
\usepackage{enumerate}
\usepackage{xspace}
\usepackage{bm}
\usepackage{comment}

\modulolinenumbers[5]

\graphicspath{{./}{figures/}{figs/}}

\journal{Journal of Computational Physics}

\newcommand{\romcap}[1]{\MakeUppercase{\romannumeral #1}}
\newcommand{\kms}{km$\cdot$s$^{-1}$}
\newcommand{\cms}{cm$\cdot$s$^{-1}$}
\newcommand{\gcmc}{g$\cdot$cm$^{-3}$}

\newcommand{\ergcm}{erg$\cdot$cm$^{-3}$}
\newcommand{\ergg}{erg$\cdot$g$^{-1}$}

\newcommand\vu{\mathbf{U}}
\newcommand\vf{\mathbf{F}}
\newcommand\va{\mathbf{A}}

\newtheorem{prop}{Proposition}
\DeclareMathOperator{\dd}{{\rm d}}

\NewDocumentCommand\pder{mmg}{\ensuremath{
		\IfNoValueTF{#3}
		{\dfrac{\partial #1}{\partial #2}}
		{\left(\dfrac{\partial #1}{\partial #2}\right)_{#3}}
	}}

\begin{document}

\begin{frontmatter}

\title{Solving the Riemann Problem for Realistic Astrophysical Fluids}

\author{Zhuo Chen}\ead{zc10@ualberta.ca}
\address{Department of Physics, University of Alberta, Edmonton, AB, T6G 2E1, Canada}
\address{Department of Physics and Astronomy, University of Rochester, Rochester, NY 14627, USA}
\address{Laboratory for Laser Energetics, University of Rochester, Rochester, NY 14623, USA}

\author{Matthew S. B. Coleman}
\address{Institute for Advanced Study, Einstein Drive, Princeton, NJ 08540, USA}
\author{Eric G. Blackman}
\address{Department of Physics and Astronomy, University of Rochester, Rochester, NY 14627, USA}
\author{Adam Frank}
\address{Department of Physics and Astronomy, University of Rochester, Rochester, NY 14627, USA}




\begin{abstract}

We present new methods to solve the Riemann problem both exactly and approximately for general equations of state (EoS) to facilitate realistic modeling and understanding of astrophysical flows. The existence and uniqueness of the new exact general EoS Riemann solution can be guaranteed if the EoS is monotone regardless of the physical validity of the EoS.
We confirm that: (1) the solution of the new exact general EoS Riemann solver and the solution of the original exact Riemann solver match when calculating perfect gas Euler equations; (2) the solution of the new Harten-Lax-van Leer-Contact (HLLC) general EoS Riemann solver and the solution of the original HLLC Riemann solver match when working with perfect gas EoS; and (3) the solution of the new HLLC general EoS Riemann solver approaches the new exact solution.
We solve the EoS with two methods, one is to interpolate 2D EoS tables by the bi-linear interpolation method, and the other is to analytically calculate thermodynamic variables at run-time. The interpolation method is more general as it can work with other monotone and realistic EoS, while the analytic EoS solver introduced here works with a relatively idealized EoS. Numerical results confirm that the accuracy of the two EoS solvers is similar. We study the efficiency of these two methods with the HLLC general EoS Riemann solver and find that analytic EoS solver is faster in the test problems. However, we point out that a combination of the two EoS solvers may become favorable in some specific problems. Throughout this research, we assume local thermal equilibrium.

\end{abstract}


\end{frontmatter}

\linenumbers

\section{Introduction}\label{sec:intro}

The need to incorporate complex equations of state (EoS) in hydrodynamics is becoming increasingly important in computational astrophysics.
The state-of-the-art for many numerically intensive research problems in such diverse sub-fields as accretion disks, binary mergers, star formation, novae, and supernovae, is approaching the state where more accurate EoS relating fluid state variables (e.g., density, pressure, and temperature) are needed. 
Although the assumption of a perfect gas EoS with a fixed constant ratio of specific heats is now prevalent in simulations with Riemann solvers, such constancy is usually not fully justified, especially where ionization or phase transitions occur (e.g., hydrogen ionization transition). In a perfect gas EoS, the specific internal energy
\begin{equation}
    \epsilon=\frac{p}{\rho(\gamma-1)},
\end{equation}
where $p,\rho$, and $\gamma$ are the pressure, density and ratio of the specific heats (i.e. $\gamma=C_{p}/C_{V}$), respectively. $C_{p}$ is the specific heat capacity at constant pressure and $C_{V}$ is the specific heat capacity at constant volume. The adiabatic sound speed $a$ of a general fluid can be calculated as
\begin{equation}
    a=\sqrt{\Gamma p/\rho},
\end{equation}
where $\Gamma=\left(\partial\ln p/\partial\ln\rho\right)_s$ and the subscript $s$ denotes that this derivative is taken at constant specific entropy.
If the fluid is a perfect gas $\Gamma=\gamma$. For a realistic gas $\gamma$ and $\Gamma$ may not be the same (see Appendix \ref{sec:thermaldynamics} for details). 

A further complication for general EoS is binding/ionization energy. For example, it takes $13.6$ eV to ionize a hydrogen atom, and the ionization happens around $T\approx 8000$ K (dependent on $\rho$), corresponding to an average thermal kinetic energy of $\approx 1.0$ eV. This makes the binding energy significantly larger than the translational energy. Similar situations occur during the disassociation of molecular hydrogen and the ionization of \ch{He} and \ch{He^+}. Neglecting these binding energies and other endothermic processes would lead to wrong gas temperature, causing significant errors in radiative transfer and chemical/nuclear reaction calculations. The hydrogen ionization instability powering outbursts observed in the accretion disks of white dwarf systems called dwarf novae \cite{L01} is one prominent example. 
Refs. \cite{H14,C16,S18} show that using an EoS, which accurately models the hydrogen ionization transition is essential for understanding dwarf novae.

The 1D general EoS hydrodynamic Euler equations that we are interested in solving are,
\begin{empheq}[left=\empheqlbrace]{align}
&\frac{\partial\rho}{\partial t}+\frac{\partial\rho u}{\partial x}=0, \\
&\frac{\partial(\rho u)}{\partial t}+\frac{\partial(\rho u^{2}+p)}{\partial x}=0, \\
&\frac{\partial E}{\partial t}+\frac{\partial(E+p)u}{\partial x} = 0,
\end{empheq}
where
\begin{align}\label{eqn:totale}
    E&=\frac{\rho u^{2}}{2}+\rho\epsilon,  \\
     &=\frac{\rho u^{2}}{2}+e_{\rm g},
\end{align}
and an EoS is required to close the system of equations
\begin{equation}\label{eqn:eos}
    \epsilon=\epsilon(\rho,p)\;\;\text{or}\;\;p=p(\rho,\epsilon),
\end{equation}
where $\rho,u,p,E,e_{\rm g}$ and $\epsilon$ are the density, velocity, pressure, total energy density, internal energy, and specific internal energy, respectively. Local thermal equilibrium (LTE) is a requirement for all hydrodynamic models, and we assume the LTE condition is satisfied throughout this paper. 
As noted before, $\Gamma=a^{2}\rho/p$ where $a$ is the adiabatic sound speed. We will use $\Gamma$ in our general EoS Riemann solvers.

Some recent work has aimed to devise efficient, high fidelity general EoS Riemann solvers \cite{toro2015,dumbser2016} with the EoS in assumed analytic form. Nonetheless, it is vital to test the accuracy of such general EoS Riemann solvers.
The idea of general EoS Riemann solver of the Harten-Lax-van Leer-Contact (HLLC) type has been mentioned by E. F. Toro in \cite{toro2013} (the last remark in Section 10.4.2, Chapter 10),
however, there has been a lack of systematic assessment of the accuracy and efficiency of the HLLC type general EoS Riemann solvers.

In this paper, we devise and examine a new efficient HLLC \cite{toro1994} general EoS Riemann solver that can handle realistic gases whose $C_{V}$, $C_{p}$ and $\Gamma$ may vary drastically when the temperature changes. The rapidly varying thermal dynamic variables may become stiff to the Euler equations. Therefore an approximate Riemann solver cannot stand alone. To justify the accuracy of the HLLC general EoS Riemann solver, we design and program a new exact general EoS Riemann solver that can work with non-convex EoS and compare the solution of the HLLC general EoS Riemann solver to the exact Riemann solver. We point out that the exact general EoS Riemann solver is the true solution with the EoS we use. Our exact general EoS Riemann solver only needs two monotone conditions,
\begin{align}
    \left(\frac{\partial p}{\partial\rho}\right)_{\epsilon}>0 \label{eqn:unique1}\\
    \left(\frac{\partial p}{\partial\epsilon}\right)_{\rho}>0  \label{eqn:unique2}
\end{align}
discussed in \cite{liu1975,smith1979}. An additional convex condition $(\partial^{2}p/\partial\rho^{2})_{s}>0$ was required by \cite{colella1985}, however we find that this is not a requirement of the new exact general EoS Riemann solver discussed in this paper (see Section \ref{sec:pnumericalmethod} for further discussion).

In Section \ref{sec:psolver}, we present the exact general EoS Riemann solver. We introduce the application of the exact general EoS Riemann solver to the Godunov scheme \cite{godunov1959} in Section \ref{sec:exactgodunov}. The novel HLLC general EoS Riemann solver will be discussed in Section \ref{sec:hllc}. Section \ref{sec:num} will focus on the numerical results generated by the exact and the HLLC general EoS Riemann solvers. In particular, we will establish the correspondence of the exact general EoS Riemann solver to the perfect gas Riemann solver, and the HLLC general EoS Riemann solver to the original HLLC Riemann solver. We obtain the solutions from the two general EoS Riemann solvers with a Godunov scheme \cite{godunov1959} and compare them to the exact solution. Section \ref{sec:num} also contains our efficiency and convergence study of the HLLC general EoS Riemann solver. We conclude in Section \ref{sec:con}.

\section{The exact general EoS Riemann solver}\label{sec:psolver}

Much of this section is a review of the seminal work \cite{colella1985}, however we utilize a more robust root finding algorithm to calculate the correct middle state pressure $p_{\rm *}$ in Section \ref{sec:pnumericalmethod}, and numerically integrate the ODEs describing the simple/rarefaction waves. We present the derivation here briefly for completeness. Readers are encouraged to read the original paper \citep{colella1985}.

\subsection{The quasi-linear hyperbolic system in non-conservative form}\label{sec:quasi-lin}

\begin{figure}
	\centering
    \includegraphics[width=0.8\textwidth]{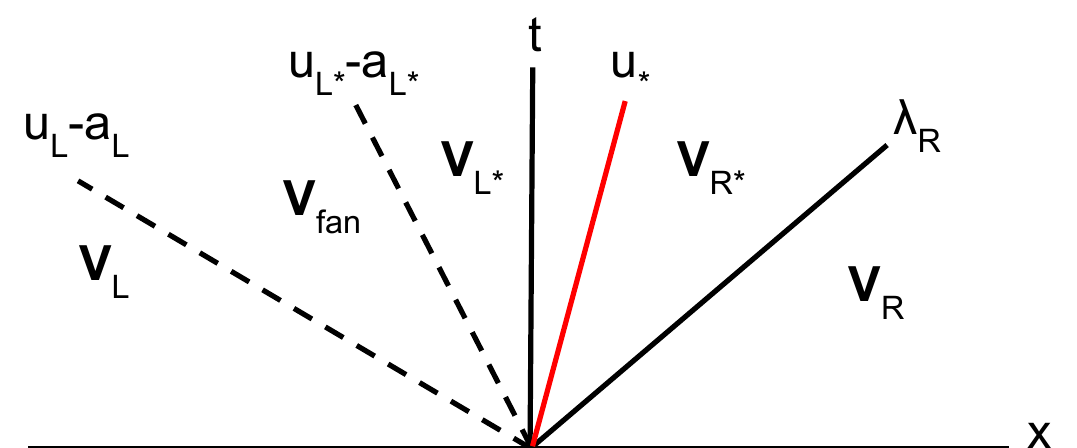}
    \caption{An illustration of the solution structure of a rightward shock tube Riemann problem. The vertical black line is the location of the interface that separates the left and right states. The region inside the two dashed lines is the rarefaction wave. The red line indicates the contact discontinuity, and it separates the left and right middle states. $\mathbf{V}_{\rm L},\mathbf{V}_{\rm R},\mathbf{V}_{\rm L*},\mathbf{V}_{\rm R*}$, and $\mathbf{V}_{\rm fan}$ are the left, right, left middle, right middle states and the state inside the rarefaction wave, respectively. $u$ and $a$ are the velocities and the adiabatic sound speed.}
    \label{RPstruct}
\end{figure}

The one-dimensional Euler equations can be written in the non-conservative form,
\begin{equation}
    \mathbf{V}_{t}+\mathbf{A}(\mathbf{V})\mathbf{V}_{x}=0,
\end{equation}
where,
\begin{equation}
    \mathbf{V}=\left[\frac{1}{\rho},u,p\right]^{T}
\end{equation}
\begin{equation}
    \mathbf{A}(\mathbf{V})=
    \begin{bmatrix}
        u & -\frac{1}{\rho} & 0 \\
        0 & u & \frac{1}{\rho} \\
        0 & \rho a^{2} & u
    \end{bmatrix},
\end{equation}
and $a$ is the adiabatic sound speed.
The (Jacobian) matrix $\mathbf{A}$ has the left and right eigenvectors ($\mathbf{l}^{1},\mathbf{r}^{1}$),($\mathbf{l}^{2},\mathbf{r}^{2}$), and ($\mathbf{l}^{3},\mathbf{r}^{3}$), which correspond to the left, middle (contact discontinuity) and right waves respectively. Figure \ref{RPstruct} shows the solution structure of a rightward shock-tube Riemann problem. 
A key condition in \cite{colella1985} is that $p_{\rm L*}=p_{\rm R*}=p_{\rm *}$ across the contact discontinuity, as previously noted by others including  \cite{hu2009}. In appendix \ref{sec:qlin}, we prove the veracity of this condition for the general EoS Euler equations.

The characteristic equations are obtained by setting $\mathbf{l}^{i}\cdot d\mathbf{V}=0$ for $i=1,2,3$:
\begin{empheq}[left=\empheqlbrace]{align}
&\mathbf{l}^{1}\cdot d\mathbf{V}=\frac{dp}{\rho a}-du=0, \label{eqn:l1}\\  
&\mathbf{l}^{2}\cdot d\mathbf{V}=\frac{dp}{(\rho a)^2}+d\left(\frac{1}{\rho}\right)=0, \label{eqn:l2}\\
&\mathbf{l}^{3}\cdot d\mathbf{V}=\frac{dp}{\rho a}+du=0. \label{eqn:l3}
\end{empheq}
For the right rarefaction wave, the solutions of $\rho,\ u$ and $p$ are calculated by Equation \ref{eqn:l1} and \ref{eqn:l2} while for the left rarefaction wave, the solutions of $\rho,\ u$ and $p$ are calculated by Equation \ref{eqn:l2} and \ref{eqn:l3}. For the left and right shock waves, the pre-shock and post-shock conditions are determined by the Rankine-Hugoniot relation:
\begin{empheq}[left=\empheqlbrace]{align}
&[u]=\mp\frac{[p]}{J_{\rm S}},  \label{eqn:rh1} \\  
&\frac{[p]}{J_{\rm S}^2}=-\left[\frac{1}{\rho}\right], \label{eqn:rh2} \\
&[\epsilon]=-\frac{p_{\rm S}+p_{\rm*}}{2}\left[\frac{1}{\rho}\right],  \label{eqn:rh3}
\end{empheq}
where $[q]=q_{\rm *}-q_{\rm S}$ and S=L,R. $J_{\rm S}=\mp\rho_{S}(\lambda_{S}-u_{S})$ is the mass flux across the shock wave. The shock front speed is expressed by $\lambda_{\rm S}=u_{\rm S}\mp\frac{J_{\rm S}}{\rho_{\rm S}}$. Substituting Equation \ref{eqn:rh3} into Equation \ref{eqn:rh2} and eliminating $\rho$, we obtain
\begin{equation}\label{eqn:JS}
    J_{\rm S}^2[\epsilon]=\frac{1}{2}[p^2].
\end{equation}
In Section \ref{sec:JS}, we will use Equation \ref{eqn:JS} to calculate $J_{\rm S}$.

\subsection{Uniqueness of solution and the p-u phase plot}

\cite{liu1975,smith1979} have proved that Equation \ref{eqn:unique1} and \ref{eqn:unique2} are sufficient but not necessary to the uniqueness of solution of Euler equation with a general EoS. We assume the gas we work with satisfy these two conditions. More in depth discussion of the EoS we use can be found in Appendix \ref{sec:eos} which contains a discussion about these two conditions.

\begin{figure}
    \centering
    \includegraphics[width=0.8\columnwidth]{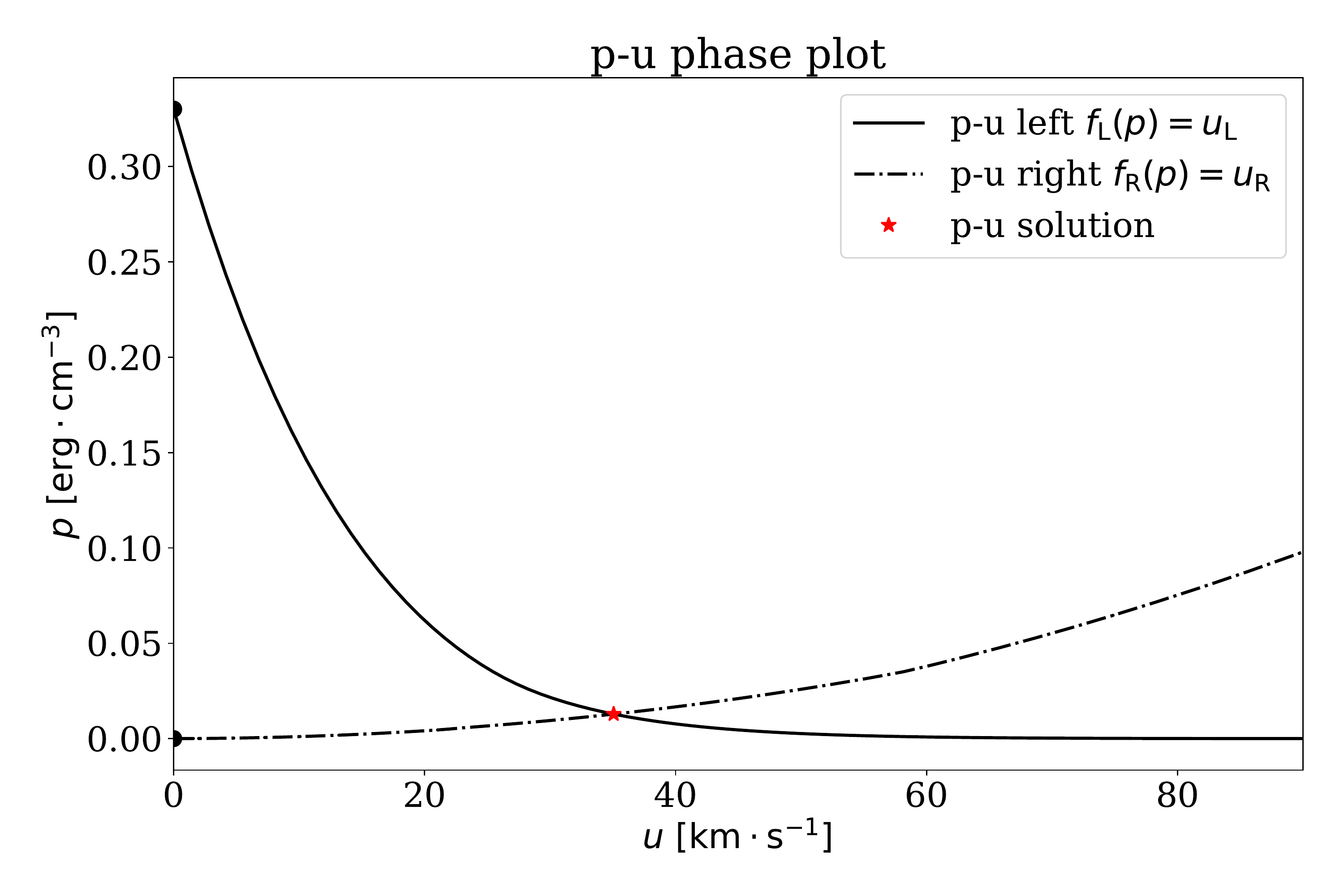}
    \caption{The p-u phase plot of a rightward shock tube test. The initial left and right states are marked with black dots.} 
    \label{fig:puphase}
\end{figure}

In Appendix \ref{sec:qlin}, Equation \ref{eqn:jumpdu} and \ref{eqn:jumpdp} prove that the contact discontinuity solution of the Euler equations with a general EoS should be $u_{\rm{L*}}=u_{\rm{R*}}=u_{\rm{*}}$ and $p_{\rm{L*}}=p_{\rm{R*}}=p_{\rm{*}}$. 
This enables us to follow \cite{CF48} and think about the solution as the intersection of curves in $p\textnormal{-}u$ phase space (See Figure \ref{fig:puphase}).
The initial state $p_{\rm S}$ and $u_{\rm S}$ is a point on the $p\textnormal{-}u$ phase plot. The rarefied or shocked state $(p_{\rm{S*}},u_{\rm{S*}})$ can be calculated by Equation \ref{eqn:l1}-\ref{eqn:rh3} and generate a continuous curve on the $p\textnormal{-}u$ phase space for each of the two values of $S$ (L, R). The following two propositions will be helpful to the understanding of the $p\textnormal{-}u$ phase plot and the uniqueness of a solution:

\begin{prop}
$dp/du<0$ for the left initial state.
\end{prop}
Proof: if $p>p_{\rm L}$, there is a shock wave on the left side, use Equation \ref{eqn:rh1} and we can get $dp/du=-J_{\rm L}<0$. If $p<=p_{\rm L}$, there is a rarefaction wave on the left side, use Equation \ref{eqn:l3} and we can get $dp/du=-1/(\rho a)<0$.

\begin{prop}
$dp/du>0$ for the right initial state.
\end{prop}
Proof: similar to \textbf{Proposition 1}.

Figure \ref{fig:puphase} shows an example $p\textnormal{-}u$ phase plot of a rightward shock with realistic EoS and a tabulated EoS solver. The initial state is the same as test 2 in Table \ref{tab:hllc}. The left and right initial states are on the p-axis in Figure \ref{fig:puphase} 
The intersection of the two curves is marked with a red star and it is the solution of $p_{*}$ and $u_{*}$. 
Let us name the curve of points that can be connected to the left state via a rarefaction wave be $f_{\rm L}(p)=u_{\rm L}$ 
and the curve of points that can be connected to the right state via a shock wave be $f_{\rm R}(p)=u_{\rm R}$.
The numerical method in Section \ref{sec:pnumericalmethod} is used to find the value of these two equations and is the main topic of the exact general EoS Riemann solver.

\subsection{Numerical method}\label{sec:pnumericalmethod}

The more robust root finding algorithm we propose is a two iterations bisection method\footnote{Other bracketing root finding techniques (e.g. the Brent-Dekker method) could also be used.} solver. In \cite{colella1985}, Secant and Newton's method were proposed to find the solution of $p_{\rm *}$. Secant and Newton's method have better convergence rate but they do not guarantee a correct solution. For example, when using these two methods to solve $f(x)=0$ and $f'(x)>0$ everywhere, if $f''(x)<0$ happens near the root, these two methods may not converge to a correct solution. In \cite{colella1985}, a convex condition, $(\partial^{2}p/\partial\rho^{2})_{s}>0$, is required to circumvent this situation. However, this condition may not be satisfied at all time in realistic gas, specifically, during the phase transition, making the secant and Newton's method incorrect. In Figure \ref{fig:cs}, we show the adiabatic sound speed $a=\left(\frac{\partial p}{\partial\rho}\right)_{s}$ of a realistic gas and emphasize the sound speed change during two phase transitions to exemplify the complexity of $a$ during phase transitions.

In the following of this section, we divide the procedure for finding the middle state pressure $p_{\rm *}$ in general EoS Riemann problem into five major steps. 

\subsubsection{Estimation of  $p_{\rm min*}$ and $p_{\rm max*}$}\label{sec:p}

There is no unique way of determining the bracketing pressure values $p_{\rm min*}$ and $p_{\rm max*}$; here we describe our procedure.
First, we need to know the maximum and minimum $\Gamma$ of the EoS table. We employ the exact perfect gas iterative Riemann solver described in Chapter 4 of \cite{toro2013} to estimate the lower and upper bound of $p_{\rm *}$. For simplicity, the solution of the exact perfect gas iterative Riemann solver is represented by a function as:
\begin{equation}\label{eqn:pexact}
    p_{\rm exact}(\mathbf{W}_{\rm L},\mathbf{W}_{\rm R},\Gamma)
\end{equation}
Where $\mathbf{W}_{\rm S}=[\rho_{\rm S},u_{\rm S},p_{\rm S}]^{T}$ is the vector of primitive (fluid) variables of the left or right state. The readers should note that Equation \ref{eqn:pexact} is actually a procedure with logical control and iteration. Equation \ref{eqn:pexact} gives the exact $p_{\rm *}$ of the middle states with the given left, right states and the $\Gamma$ for perfect gas. We are slightly abusing this procedure here because $\Gamma\ne\gamma$ for realistic gas. However, we only use them for an initial guess of the lower and upper bound in which $p_{\rm *}$ may reside in and it turns out to be useful.
The lower and upper bound of $p_{\rm *}$ are estimated by,
\begin{align}\label{eqn:pmin}
    p_{\rm min*}&=p_{\rm exact}(\mathbf{W}_{\rm L},\mathbf{W}_{\rm R},\Gamma_{\rm max})/2\\
\label{eqn:pmax}
    p_{\rm max*}&=p_{\rm exact}(\mathbf{W}_{\rm L},\mathbf{W}_{\rm R},\Gamma_{\rm min})\times2
\end{align}
We multiply or divide the guessed pressure by a factor of two to ensure that the solution is indeed bracketed.

\subsubsection{Estimate  $J_{\rm S,min}$ and $J_{\rm S,max}$ for $p_{\rm *}>p_{\rm S}$ (shock present)}\label{sec:JS}

When $p_{\rm *}>p_{\rm S}$, there is a left or right shock. For each estimate of $p_{\rm *}>p_{\rm S}$ in the range $[p_{\rm min*},p_{\rm max*}]$ during the $p_{*}$ bisection iteration (see Section \ref{sec:bisection}), we estimate the corresponding range of $J_{\rm S}$ ($[J_{\rm S,min}$, $J_{\rm S,max}]$) and initiate the second bisection loop. We use the chosen $p_{\rm *}$ to estimate  $J_{\rm S,min}$ as
\begin{equation}
	J_{\rm S,min}=0,
\end{equation}
since the mass flux could be close to $0$ according to the definition of $J_{\rm S}$, and
\begin{equation}
	J_{\rm S,max}=10\times\rho_{\rm \bar{S}}(|v_{\rm \bar{S}}|+a_{\rm \bar{S}}),
\end{equation}
where $\bar{\rm S}$ is the opposite state of $\rm S$ and $a_{\rm \bar{S}}$ is the adiabatic sound speed of the $\bar{S}$ state.
The  choice of a large $J_{\rm S,max}$  is  heuristic because we are presently interested  only in the correctness of the exact general EoS Riemann solver, not its efficiency. This means that we only require that the root ($J_{\rm S}$) is bracketed. Practically, one may not be able to apply the exact general EoS Riemann solver to 2D and 3D calculations even with an optimized algorithm.

\subsubsection{Using the bisection method to determine  $p_{\rm *}$ and $J_{\rm S}$}\label{sec:bisection}

In this section, we show that our bisection method ($J_{\rm S}$ iteration) guarantees to find one and only one solution of $J_{\rm S}$. 

Let $u_{\rm L*}$ and $u_{\rm R*}$ be the velocity at the left and right side of the contact discontinuity calculated with any given $p_{\rm *}$. If $p_{\rm *}$ exists within the range $[p_{\rm min*},p_{\rm max*}]$, then the expression $f(p_{\rm *})=u_{\rm L*}-u_{\rm R*}$, should have different signs at $p_{\rm min*}$ and $p_{\rm max*}$, otherwise the solution has not been properly bracketed. $\Delta u$ can be defined as,
\begin{equation}
    \Delta u(p_{\rm *})=f_{\rm L}(p_{\rm *})-f_{\rm R}(p_{\rm *})
\end{equation}
where $f_{\rm L}$ corresponds to the curve of left state in the $p\textnormal{-}u$ phase space and $f_{\rm R}$ corresponds to the curve of right state in the $p\textnormal{-}u$ phase space (Figure \ref{fig:puphase}). Put another way, inserting the correct $p_{\rm *}$ will yield $u_{\rm L*}=u_{\rm R*}$ and $\Delta u(p_{\rm *})=0$. We find the correct $p_{\rm *}$ in the range  $[p_{\rm min*},p_{\rm max*}]$ by first picking a $p_{\rm *}$ in the range  $[p_{\rm min*},p_{\rm max*}]$.

If $p_{\rm *}>p_{\rm S}$, then there is a shock wave on the $S$ side. Find the correct $J_{\rm S}$ in the range of $[J_{\rm S,min},J_{\rm S,max}]$ using the bisection method. The correct $J_{\rm S}$ should satisfy Equations \ref{eqn:eos}, \ref{eqn:rh2}, and \ref{eqn:JS} simultaneously. For convenience, we list and rearrange the equations here again,
\begin{empheq}[left=\empheqlbrace]{align}
    \epsilon&=\epsilon(\rho,p), \label{eqn:eoscopy}\\
    \frac{1}{\rho_{\rm S*}}&=\frac{1}{\rho_{\rm S}}-\frac{[p]}{J_{\rm S}^2}, \label{eqn:rho*copy}\\
    J_{\rm S}^2[\epsilon]&=\frac{1}{2}[p^2]. \label{eqn:JScopy}
\end{empheq}
Equation \ref{eqn:rho*copy} relates $\rho_{\rm S*}$ to $J_{\rm S}^{2}$. Substitute Equations \ref{eqn:eoscopy} and \ref{eqn:rho*copy} into Equation \ref{eqn:JScopy}, to obtain a single equation for $J_{\rm S}^{2}$. For convenience, let us define  $W\equiv J_{\rm S}^{2}\ge0$. Then 
\begin{equation}\label{eqn:JSbisection}
	W\left(\epsilon_{\rm S*}-\epsilon\right)=\frac{p_{\rm *}^{2}-p^{2}}{2},
\end{equation}
where $\epsilon_{\rm S*}=\epsilon_{\rm *}(W)$ is implicitly a function of $W$ (via Equations~\ref{eqn:eoscopy}-\ref{eqn:JScopy}).
The right hand side of Equation \ref{eqn:JSbisection} is a positive constant in the iteration of $J_{\rm S}$. The derivative of the left hand side is 
\begin{equation}
	\frac{d\text{LHS}}{dW}=\epsilon_{\rm *}(W)-\epsilon+W\left(\frac{\partial\epsilon_{\rm S*}}{\partial\rho_{\rm S*}}\right)_{p_{\rm *}}\frac{d\rho_{\rm S*}}{dW},
\end{equation}
from Equation \ref{eqn:rho*copy}
\begin{align}
	\frac{d\rho_{\rm S*}}{dW}=-\rho_{\rm S*}^2\frac{p_{\rm *}-p}{W^2}<0,
\end{align}
and from Equation \ref{eqn:unique1} and \ref{eqn:unique2},
\begin{align}
	\left(\frac{\partial\epsilon}{\partial\rho}\right)_{p}=-\left(\frac{\partial\epsilon}{\partial p}\right)_{\rho}\bigg/\left(\frac{\partial\rho}{\partial p}\right)_{\epsilon}=\epsilon_{\rho}\equiv\epsilon_{\rho_{\rm S*}}<0.
\end{align}
Thus, $d\text{LHS}/dW\ge0$ and only equals 0 when $W=J_{\rm S}^{2}=0$, $\text{LHS}=0$ when $W=J_{\rm S}^{2}=0$, so there is exactly one solution of $W$ to Equation \ref{eqn:JSbisection}. In Section \ref{sec:JS}, we have chosen the $J_{\rm S,max}$ to be large enough to ensure the solution $J_{\rm S}\le J_{\rm S,max}$.
    

To apply the bisection root finding algorithm to Equation \ref{eqn:JSbisection}, we  pick a $J_{\rm S}$ in the range of $[J_{\rm S,min},J_{\rm S,max}]$. We then calculate $\rho_{\rm *}$ using Equation \ref{eqn:rho*copy} and calculate $\epsilon_{\rm *}$ using Equation \ref{eqn:eoscopy} for the choice of  $J_{\rm S}$ and substitute these results into Equation \ref{eqn:JSbisection}. If LHS$\ne$RHS, we must iterate the bounds of $J_{\rm S}$ until the bracketing range narrows in on a solution. A solution of $J_{\rm S}$ is thought to be found if:
\begin{equation}
    J_{\rm S,max}^{2}-J_{\rm S,min}^{2}<2\delta_{\rm{machine}}
\end{equation}
where $J_{\rm S,max}$ and $J_{\rm S,min}$ are the updated upper and lower bound of the bracket. $\delta_{\rm{machine}}$ is the machine precision unit.

After a proper $J_{\rm S}$ has been found, $u_{\rm{S*}}$ can be calculated by Equation \ref{eqn:rh1}.
    
If $p_{\rm *}\le p_{\rm S}$, there is a rarefaction wave on the $S$ side. We then simply solve the ODEs \ref{eqn:l1} and \ref{eqn:l2} for a right rarefaction wave and \ref{eqn:l2} and \ref{eqn:l3} for a left rarefaction wave. The definite integration starts at $p_{\rm S}$ and ends at $p_{\rm *}$. We then calculate  $u_{\rm S*}$.
    
After obtaining $u_{\rm L*}$ and $u_{\rm R*}$ with the given $p_{\rm *}$, we compare $u_{\rm L*}-u_{\rm R*}$ to the value of 0, update $p_{\rm min*}$ or $p_{\rm max*}$ and increment $p_{\rm *}$ according to bisection method.

The iteration is stopped when $p_{\rm min*}$ and $p_{\rm max*}$ are sufficiently close. In practice, we define ``sufficiently close" as,
\begin{equation}\label{eqn:delta1}
    p_{\rm max*}-p_{\rm min*}<2\delta_{\rm{machine}},
\end{equation}
When condition \ref{eqn:delta1} is satisfied, $\left|u_{\rm{R*}}-u_{\rm{L*}}\right|$ may not be less than $2\delta_{\rm{machine}}$. Different problems may also give different gaps. This is because $p_{*}$ and $u_{*}$ are in general converging at different rate, it is impractical to find a constant $\delta$ and force their gaps to become smaller than $\delta$ at the same iteration. In addition, the nature of bisection method does not allow further numerical calculation if condition \ref{eqn:delta1} is satisfied therefore we would just stick to this convergence criterion. Calculation shows that $\left|u_{\rm{R*}}-u_{\rm{L*}}\right|<10^{-5}$ \cms\ for all of our tests and the gap decreases as $\left|p_{\rm max*}-p_{\rm min*}\right|$ decreases.

\subsubsection{Calculation of  $\lambda_{\rm S}$ and $u_{\rm *}$}\label{sec:wavespeeds}
To calculate the three wave speeds discussed in Section~\ref{sec:quasi-lin} for a shock wave, we need to utilize the $J_{\rm S}$ computed in the previous section:
\begin{align}
    \lambda_{\rm S}&=u_{\rm S}\mp\frac{J_{\rm S}}{\rho_{\rm S}}\\
    u_{\rm *}&=u_{\rm S}\mp\frac{p_{\rm *}-p_{\rm S}}{J_{\rm S}}.
\end{align}
If there is a rarefaction wave, then
\begin{equation}
    \lambda_{\rm S}=u_{\rm S}\mp a_{\rm S},
\end{equation}
where $a_{\rm S}$ is the adiabatic sound speed of the S state. Here $u_{\rm *}$ can be calculated by solving ODEs \ref{eqn:l1} and \ref{eqn:l2} for the right rarefaction wave and ODEs \ref{eqn:l2} and \ref{eqn:l3} for the left rarefaction wave, where the bounds of the integration are $p_{\rm S}$ and $p_{\rm *}$. In any given Riemann problem, the value $u_{\rm *}$ calculated by the left and right waves must be equal (as proved in Appendix \ref{sec:qlin}), otherwise the previous steps of our solution procedure have not been followed properly (e.g. finding $p_{\rm *}$ or $J_{\rm S}$).

\subsubsection{Sample results}\label{sec:sampleresults}

\begin{figure}
	\centering
    \includegraphics[width=0.9\textwidth]{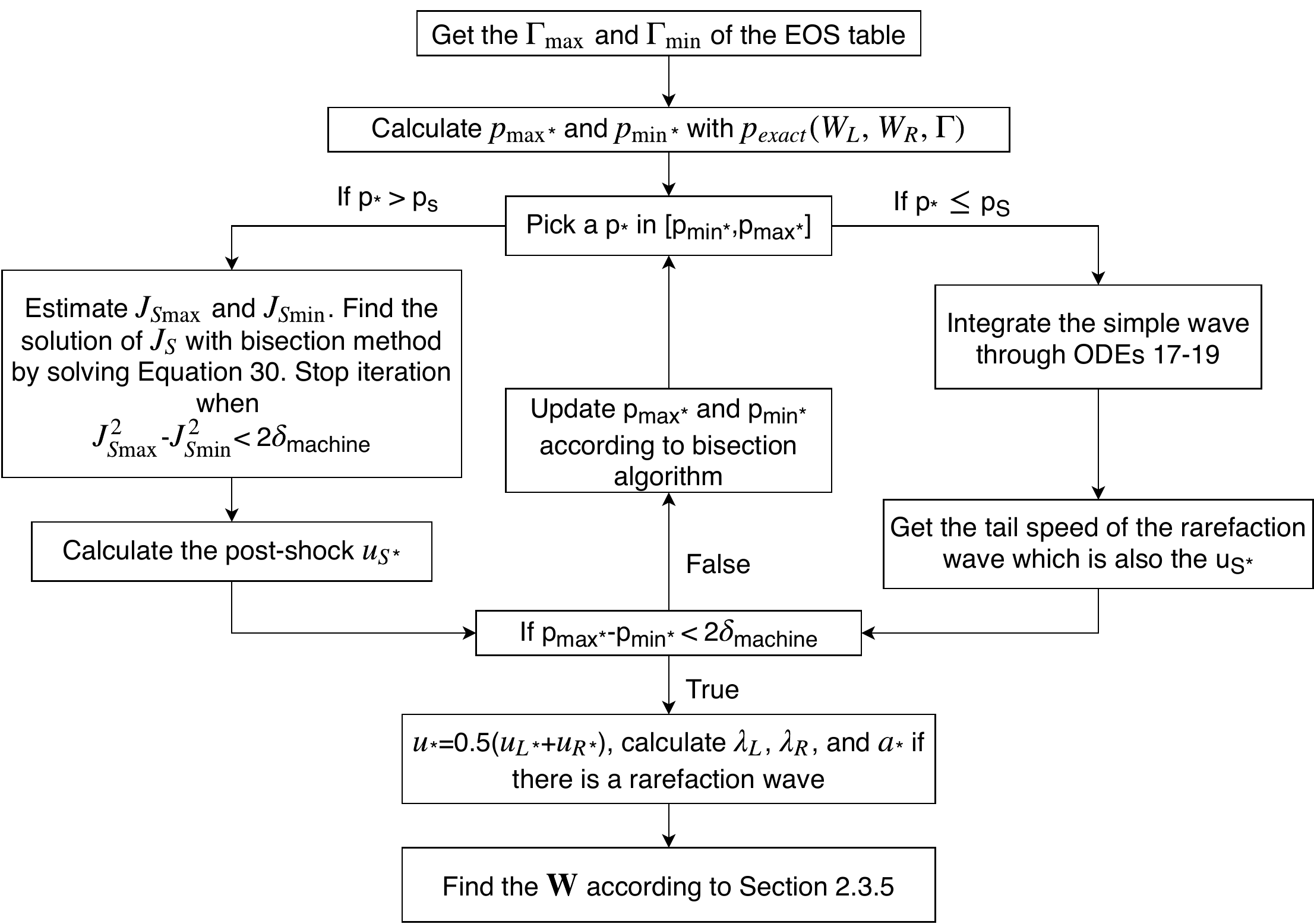}
    \caption{The flow chart of the main steps of the exact general EoS Riemann solver.}
    \label{fig:flowchart}
\end{figure}

The solution ($\mathbf{W}(x,T)$) of a Riemann problem could be one of the three types: (i) two shock waves; (ii) two rarefaction waves: (iii) and one shock wave with one rarefaction wave. 
As before the left wave speed is $\lambda_{\rm L}$ and the right wave speed is $\lambda_{\rm R}$. The ratio of coordinate $x$ to given time $t$ is related to the different waves and state vectors $\mathbf{W}(x,T)=\mathbf{W}$ as follows:\\
For the left shock wave,
\begin{itemize}
    \item If $x/t<\lambda_{\rm L}$, then $\mathbf{W}=\mathbf{W}_{\rm L}$
    \item If $\lambda_{\rm L}\le x/t<u_{\rm *}$, then $\mathbf{W}=\mathbf{W}_{\rm L*}$
\end{itemize}
For the left rarefaction wave,
\begin{itemize}
    \item If $x/t<\lambda_{\rm L}$, then $\mathbf{W}=\mathbf{W}_{\rm L}$
    \item If $\lambda_{\rm L}\le x/t<u_{\rm *}-a_{\rm *}$, integrate the ODEs \ref{eqn:l2} and \ref{eqn:l3} until $x/t=u-a$ to get the correct $\mathbf{W}$ in the fan region.
    \item If $u_{\rm *}-a_{\rm *}\le x/t<u_{\rm *}$, then $\mathbf{W}=\mathbf{W}_{\rm L*}$
\end{itemize}
For the right shock wave,
\begin{itemize}
    \item If $\lambda_{\rm R}<x/t$, then $\mathbf{W}=\mathbf{W}_{\rm R}$
    \item If $u_{\rm *}< x/t\le\lambda_{\rm R}$, then $\mathbf{W}=\mathbf{W}_{\rm R*}$
\end{itemize}
For the right rarefaction wave,
\begin{itemize}
    \item If $\lambda_{\rm R}<x/t$, then $\mathbf{W}=\mathbf{W}_{\rm R}$
    \item If $u_{\rm *}+a_{\rm *}\le x/t<\lambda_{\rm R}$, integrate \ref{eqn:l1} and \ref{eqn:l2} until $x/t=u+a$ to get the correct $\mathbf{W}$ in the fan region.
    \item If $u_{\rm *}\le x/t<u_{\rm *}+a_{\rm *}$, then $\mathbf{W}=\mathbf{W}_{\rm R*}$
\end{itemize}
We present a flow chart of the exact general EoS Riemann solver in Figure \ref{fig:flowchart} to make it easier to follow.

\subsection{Tabulated EoS solver for the exact general EoS Riemann solver}\label{sec:exacttabulated}

In the exact general EoS Riemann solver, we constantly need to solve
\begin{align}
    &a(\rho,p)       \label{eqn:tabulateda}\\
    &\epsilon(\rho,p)   \label{eqn:tabulatedepsilon}
\end{align}
in Equation \ref{eqn:l1}-\ref{eqn:rh3}. Therefore, we prepare tables with of $\log a(\log\rho,\log p)$ and $\log\epsilon(\log\rho,\log p)$ and use bi-linear interpolation method to interpolate the tables to get the value of $a$ and $\epsilon$. Bi-linear interpolation method is easy to implement and also preserve the monotonicity of the table. The tables we use are rectangular and the domain of the tables are $\log\rho\in[-15,-10]$ \gcmc and $\log p\in[-5.38,2.62]$ \ergcm. The resolution of the domain is $501\times701$ including the endpoints.

We can calculate $\epsilon,\ T$ and $\mu$ by interpolating $\log\epsilon(\log\rho,\log p)$, $\log T(\log\rho,\log p)$, and $\mu(\log\rho,\log p)$ 
tables with bi-linear interpolation method. The number density $n_{i}$ of all species can be calculated after getting the value of $\mu$.

\section{Applying the exact general EoS Riemann solver to the Godunov scheme}\label{sec:exactgodunov}

In this section, we discuss how to apply the exact general EoS Riemann solver to the Godunov scheme to first order. For an introduction to the Godunov scheme, we refer readers to standard computational fluid mechanics books, for example \cite{leveque2002,toro2013}.
The quantities that are needed in the Godunov scheme are the flux and the fastest wave speed at the interface. In Section \ref{sec:psolver}, we have solved the exact general EoS Riemann problem. We are able to calculate the $\mathbf{W}$ at any point and get the nonlinear wave speeds. We can easily compare the wave speeds and pick the fastest one. To calculate the flux, we should set $dx/dt=x/t=0$ in Section \ref{sec:sampleresults} and calculate the corresponding $\mathbf{W}$ and the flux through the flux function
\begin{align}\label{eqn:fluxfunction}
\vf=
\begin{pmatrix}
\rho u\\
\rho u^2+p\\
u(E+p)
\end{pmatrix}.
\end{align}

\section{The Harten-Lax-van Leer-Contact (HLLC) general EoS Riemann solver}\label{sec:hllc}

\subsection{A brief review of the original HLLC Riemann solver}

\begin{figure}
    \centering
    \includegraphics[width=0.8\textwidth]{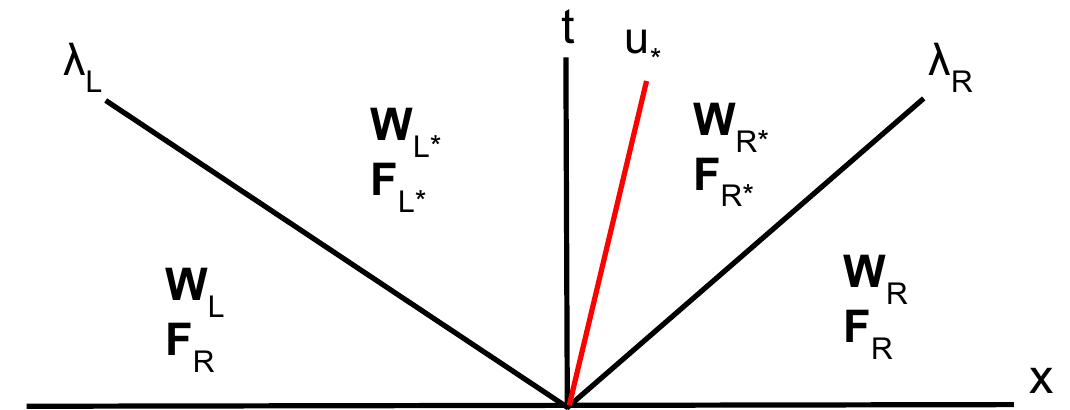}
    \caption{the solution structure of a classic HLLC Riemann solver consists of a left wave $\lambda_{\rm L}$, a right wave $\lambda_{\rm R}$ and a contact wave $u_{\rm *}$. The four primitive states divided by the three waves are $\mathbf{W}_{\rm L},\mathbf{W}_{\rm L*},\mathbf{W}_{\rm R*}$ and $\mathbf{W}_{\rm R}$. $\mathbf{F}_{\rm S}=[\rho_{\rm S}u_{\rm S},\rho_{\rm S}u_{\rm S}^{2}+p_{\rm S},u_{\rm S}(E_{\rm S}+p_{\rm S})]^{T}$ is the corresponding flux associated with the primitive state.}
    \label{fig:hllc}
\end{figure}

We assume that readers are familiar with the original HLLC Riemann solver \citep{toro1994}
and only briefly list the main steps here. The original HLLC Riemann solver differs from the HLL \citep{harten1983} Riemann solver by dividing the middle region into two states separated by the middle wave (i.e. the contact discontinuity). Figure \ref{fig:hllc} shows a solution structure of the original HLLC Riemann solver, where $\lambda_{\rm L}$ and $\lambda_{\rm R}$ are the speeds of the left and the right waves, and $u_{\rm *}$ is the middle wave speed. The main difference between the exact and HLLC Riemann Solvers is that HLLC assumes that the left and right waves are always shocks.
The wave speeds $\lambda_{\rm L}$ and $\lambda_{\rm R}$ need to be calculated before calculating the middle wave speed:
\begin{align}\label{eqn:hllcsl}
    \lambda_{\rm L}&=u_{\rm L}-a_{\rm L}q_{\rm L},\\
\label{eqn:hllcsr}
    \lambda_{\rm R}&=u_{\rm R}+a_{\rm R}q_{\rm R},
\end{align}
where
\begin{equation}\label{eqn:qratio}
q_{\rm S}=
\begin{cases}
  1, & p{*}\le p_{\rm S}
\\
  \sqrt{1+\frac{\Gamma_{\rm S}+1}{2\Gamma_{\rm S}}(\frac{p_{\rm *}}{p_{\rm S}}-1)}, & p{*}>p_{\rm S}
\end{cases}
\end{equation}

Equation \ref{eqn:hllcsl}-\ref{eqn:qratio} is referred as the ``hybrid estimate" in \cite{toro1994}. We briefly describe the estimation of $p_{*}$ in Section \ref{sec:estimatep}.

Additional constraints are placed by the the Rankine-Hugoniot conditions (Equations~\ref{eqn:rh1}-\ref{eqn:rh3}) which can be rewritten as follows (Section 10.4.1, Chapter 10 of \cite{toro2013}):
\begin{empheq}[left=\empheqlbrace]{align}\label{eqn:fl}
\mathbf{F}_{\rm L*}&=\mathbf{F}_{\rm L}+\lambda_{\rm L}(\mathbf{U}_{\rm L*}-\mathbf{U}_{\rm L})\\\label{eqn:fr1}
\mathbf{F}_{\rm R*}&=\mathbf{F}_{\rm L*}+u_{\rm *}(\mathbf{U}_{\rm R*}-\mathbf{U}_{\rm L*})\\\label{eqn:fr2}
\mathbf{F}_{\rm R*}&=\mathbf{F}_{\rm R}+\lambda_{\rm R}(\mathbf{U}_{\rm R*}-\mathbf{U}_{\rm R}),
\end{empheq}
where $\mathbf{U}_{\rm S}=[\rho_{\rm S},\rho_{\rm S}u_{\rm S},E_{\rm S}]^{T}$ are the conservative quantities, and $\lambda_{\rm L},\lambda_{\rm R}$ and $u_{\rm *}$ are three waves speeds as before. 
The original HLLC Riemann solver also reintroduces the contact discontinuity
\begin{align}
    p_{\rm L*}&=p_{\rm R*}=p_{\rm *}\\
    u_{\rm L*}&=u_{\rm R*}=u_{\rm *}
\end{align}
to HLL type Riemann solvers. With the addition of these constraints, the middle wave speed can be calculated
\begin{equation}\label{eqn:u*}
    u_{\rm *}=\frac{p_{\rm R}-p_{\rm L}+\rho_{\rm L}u_{\rm L}(\lambda_{\rm L}-u_{\rm L})-\rho_{\rm R}u_{\rm R}(\lambda_{\rm R}-u_{\rm R})}{\rho_{\rm L}(\lambda_{\rm L}-u_{\rm L})-\rho_{\rm R}(\lambda_{\rm R}-u_{\rm R})},
\end{equation}
enabling the computation of the two middle states,
\begin{equation}\label{eqn:middlestates}
    \mathbf{U}_{\rm S*}=\rho_{\rm S}\left(\frac{\lambda_{\rm S}-u_{\rm S}}{\lambda_{\rm S}-u_{\rm *}}\right)
    \begin{bmatrix}
        1 \\
        u_{\rm *} \\
        \frac{E_{\rm S}}{\rho_{\rm S}}+(u_{\rm *}-u_{\rm S})\left[u_{\rm *}+\frac{p_{\rm S}}{\rho_{\rm S}(\lambda_{\rm S}-u_{\rm S})}\right]
    \end{bmatrix}.
\end{equation}

\subsection{Algorithms of the HLLC general EoS Riemann solver}

Ideally, the HLLC general EoS Riemann solver should reduce to the original HLLC Riemann solver when supplied with EoS tables of perfect gas. 
Consequently, we attempted to minimize the changes when expanding the HLLC method to general EoS.
For completeness, we also outline the whole procedure here.

\subsubsection{Estimate $p_{*}$}\label{sec:estimatep}

Unlike the perfect gas, realistic gas has variable $\Gamma$. When the $\Gamma_{\rm L}$ of the left and $\Gamma_{\rm R}$ of the right state are very different, it may be natural to use one of the $\Gamma$. We have experimented with $\Gamma=\min{(\Gamma_{\rm L},\Gamma_{\rm R})}$, $\Gamma=\max{(\Gamma_{\rm L},\Gamma_{\rm R})}$, $\Gamma=(\Gamma_{\rm L}+\Gamma_{\rm R})/2$, the upwind or downwind $\Gamma$, and both $\Gamma_{\rm L}$ and $\Gamma_{\rm R}$. We find the difference in them is not big in terms of the solution of $\rho,\ u,\ p,\ T,\ e_{\rm g}$, etc. The reason for this could be that the solution of $\Gamma$ is already different from the exact solution (see Figure \ref{fig:eossolver1}-\ref{fig:eossolver4}) because $\Gamma$ is very sensitive to $\rho$ and $T$. If $\Gamma$ is not accurate, any manipulation of $\Gamma$ would not be effective. In this paper, we use both $\Gamma$ to calculate $p_{*}$. The algorithm that we use to calculate $p_{*}$ is a small variation of the adaptive non-iterative Riemann solver in Section 9.5.2, Chapter 9 of \cite{toro2013}. It consists of three sub-solvers: the primitive variable, two-rarefaction, and two-shock Riemann solvers. We substitute all the sound speed with the adiabatic sound speed calculated from the EoS solvers, which means we are using both $\Gamma_{\rm L}$ and $\Gamma_{\rm R}$. In the two-rarefaction Riemann solver, one $\Gamma$ is still needed, in this situation, we use $\Gamma=\min(\Gamma_{\rm L},\Gamma_{\rm R})$.

In the above process of estimation, the only possible values are $\Gamma_{\rm L}$ and $\Gamma_{\rm R}$. Therefore, in the case of a perfect gas, the estimated $\Gamma$ will always be the constant value (i.e., $\Gamma=\Gamma_{\rm L}=\Gamma_{\rm R}=\gamma$). This guarantees that the HLLC general EoS Riemann solver can be reduced to the original HLLC Riemann solver when fed with an EoS table with constant $\Gamma$. We will demonstrate this with two numerical examples in Section \ref{sec:hllcvshllc}.

We choose this adaptive non-iterative Riemann solver because it is simple and fast. We would admit this choice is quite heuristic at this stage in the context of general EoS. Other choices of Riemann solver are also available \citep{Batten1997}. We find that using the Roe average Riemann solver \cite{Batten1997} gives similar results to the one introduced here, but it is slower in our implementation as we need to solve the EoS to get the sound speed of the middle state.

\subsubsection{Calculate $\lambda_{\rm S}$, $u_{\rm *}$, $\mathbf{U}_{\rm S*}$ and the flux}

\begin{figure}
    \centering
    \includegraphics[width=0.8\textwidth]{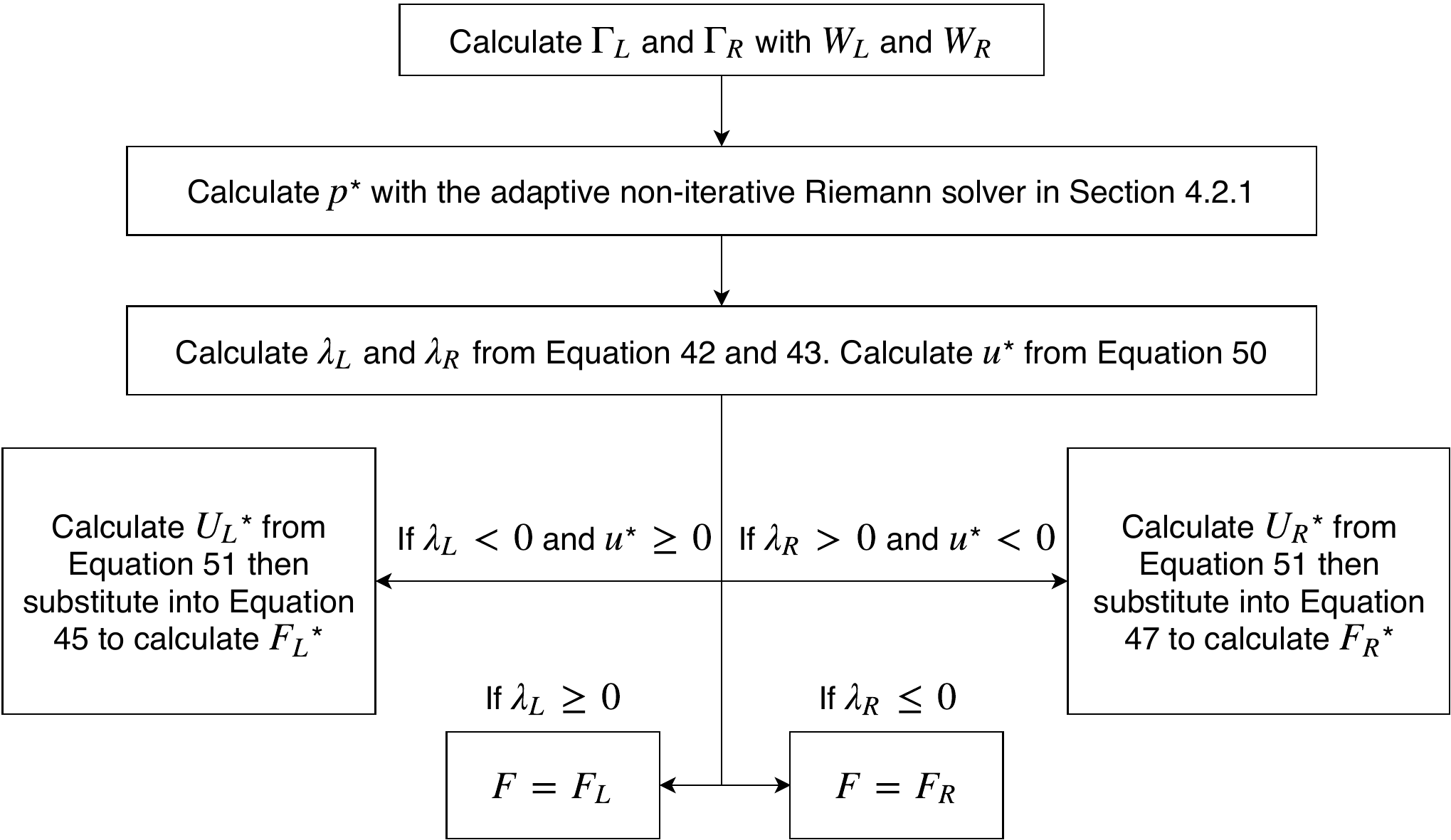}
    \caption{The flow chart of the main steps of the HLLC general EoS Riemann solver.}
    \label{fig:eoshllc}
\end{figure}

\begin{itemize}
    \item Calculate $p_{\rm *}$ with the procedure described in Section \ref{sec:estimatep}.
    \item Calculate $\lambda_{\rm L}$ or $\lambda_{\rm R}$ from Equation \ref{eqn:hllcsl} and \ref{eqn:hllcsr}. Use the left and right adiabatic sound speeds that are interpolated from the EoS tables or calculated by the analytic EoS solver.
    \item Calculate $u_{\rm *}$ from Equation \ref{eqn:u*} with the given $\lambda_{\rm L}$ and $\lambda_{\rm R}$.
    \item The $U_{\rm S*}$ can be calculated by Equation \ref{eqn:middlestates}.
    \item Calculate the intermediate flux from Equation \ref{eqn:fl} or \ref{eqn:fr2} with all other calculated quantities.
\end{itemize}
Figure \ref{fig:eoshllc} shows the main steps to get the flux at the interface of the HLLC general EoS Riemann solver.
After getting the flux, it is straightforward to apply it to any conservative schemes (for example, Godunov scheme) and calculate the conserved quantities. In designing the algorithm of the HLLC general EoS Riemann solver, we intentional keep the change to a low level to achieve a faster speed. The changes in our procedure compared to the original HLLC Riemann solver include an estimation the $\Gamma$ before the calculations and interpolation of the adiabatic sound speed. These relatively small changes result in numerically accurate solutions as we show if the following section.

\subsection{EoS solvers for the HLLC general EoS Riemann solver}\label{sec:hllceossolver}

We have implemented two types of EoS solvers for the HLLC type of general EoS Riemann solver. The tabulated EoS solver uses the interpolation method described in Section \ref{sec:exacttabulated}. In addition, we need to be able to calculate
\begin{align}
    &\Gamma(\rho,p),   \label{eqn:tabulatedgamma}\\
    &p(\rho,\epsilon)       \label{eqn:tabulatedp}.
\end{align}
Equation \ref{eqn:tabulatedgamma} is needed in calculating the left and right adiabatic sound speed. Equation \ref{eqn:tabulatedp} will be used to convert the conservative quantities to the primitive quantities. We use bi-linear interpolation method to interpolate $\Gamma(\log\rho,\log p)$ and $\log p(\log\rho,\log\epsilon)$. The domain of the table $(\log\rho,\log\epsilon)$ is $\log\rho\in[-15,-10]$ \gcmc and $\log\epsilon\in[9.79,13.33]$ \ergg with a resolution of $501\times701$ including endpoints.

The analytic EoS solver is discuss in detail in Appendix \ref{sec:eos}.

\section{Numerical results}\label{sec:num}

\subsection{The correspondence between the exact general EoS Riemann solver and the exact perfect gas Riemann solver}\label{sec:exactvsexact}

\begin{figure}
    \centering
    \includegraphics[width=0.495\textwidth]{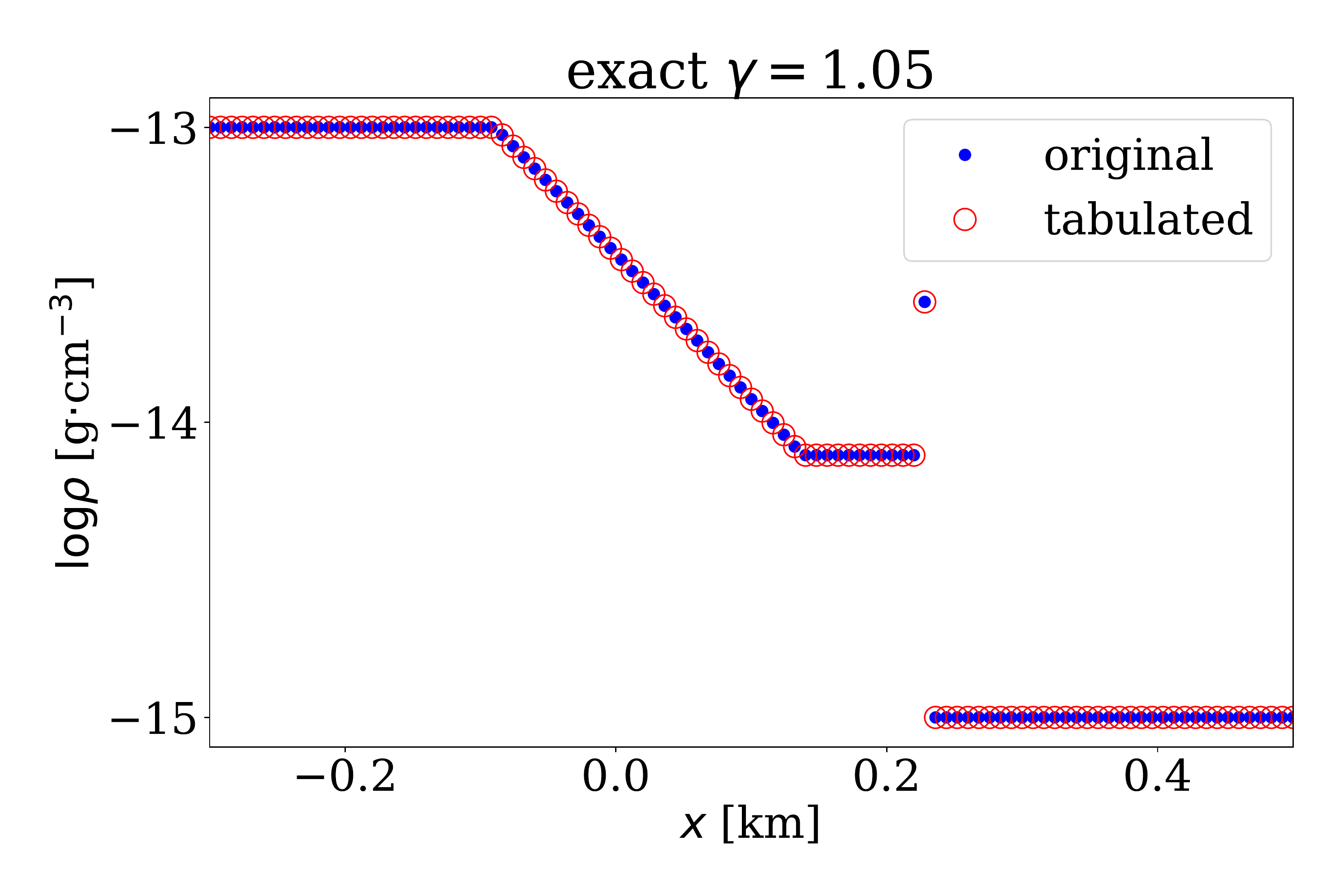}
    \includegraphics[width=0.495\textwidth]{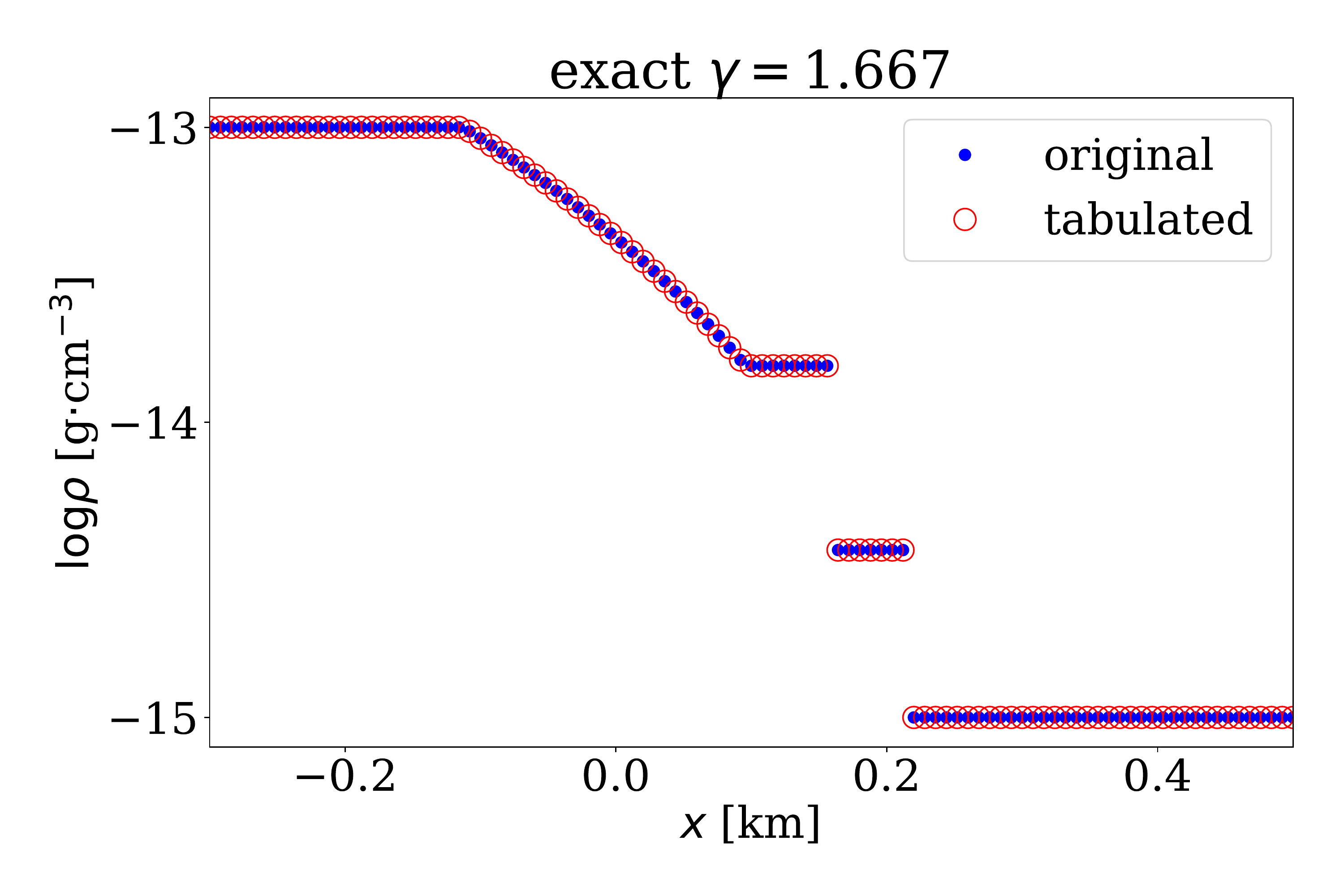}
    \caption{The solutions of density of the exact general EoS and the exact perfect gas Riemann solver at $t_{\rm{max}}=0.02$s. $\gamma=\Gamma=1.05$ and $1.667$, respectively.}
    \label{fig:exactvsexact1}
\end{figure}

The perfect gas Riemann solver should be a special case of the general EoS Riemann solver. When using the EoS tables of a perfect gas, the general EoS Riemann solver should give the same solution as the original exact ideal gas Riemann solver. To test this aspect, we create EoS tables of two kinds of perfect gas differ in $\gamma=\Gamma=1.05,1.667$, but the same mean molecular weight $\mu=1.3$. The initial left and right states are listed in Table \ref{tab:lrstate}. The left and right states are separated by origin. We solve the EoS by interpolation method described in Section \ref{sec:exacttabulated}.
The results of both exact Riemann solvers are shown in Figure \ref{fig:exactvsexact1}. Red circles and blue dots indicate the solution from the exact general EoS Riemann solver and the solution from the exact perfect gas Riemann solver, respectively. The solutions from the two exact Riemann solvers are very close, confirming that the exact general EoS Riemann solver can be reduced to an exact perfect gas Riemann solver when a perfect gas EoS is used. Figure \ref{fig:exactvsexact1} also shows that low $\gamma$ perfect gas has much higher compressibility compared to the high $\gamma$ perfect gas.

\begin{table}
    \centering
    \begin{tabular}{|c|c|c|c|c|c|c|c|}\hline
      ID  &  $t_{\rm{max}}$  &   $\rho_{\rm L}$  &  $u_{\rm L}$   &  $T_{\rm L}$     &  $\rho_{\rm R}$  &  $u_{\rm R}$   &  $T_{\rm R}$     \\
         &  [s]  &   [\gcmc]    &   [\kms]  &   [K]    &   [\gcmc]   &   [\kms]  &   [K]     \\ \hline
     1  &  0.02   &   $10^{-13}$  &  $0$   &  $3000$    &  $10^{-15}$  &  $0$  &   $300$     \\ \hline
    \end{tabular}
    \caption{Test 1 is a shock tube test. We choose the ratio of specific heat $\gamma=1.05$ and $1.667$ while keep the mean atomic weight $\mu=1.3$. The density, velocity and temperature of the left state are listed from the second column to the fourth column while the rightest three columns list the variables of the right state.}
    \label{tab:lrstate}
\end{table}

\subsection{The correspondence between the HLLC general EoS Riemann solver and the original HLLC Riemann solver}\label{sec:hllcvshllc}

\begin{figure}
    \centering
    \includegraphics[width=0.495\textwidth]{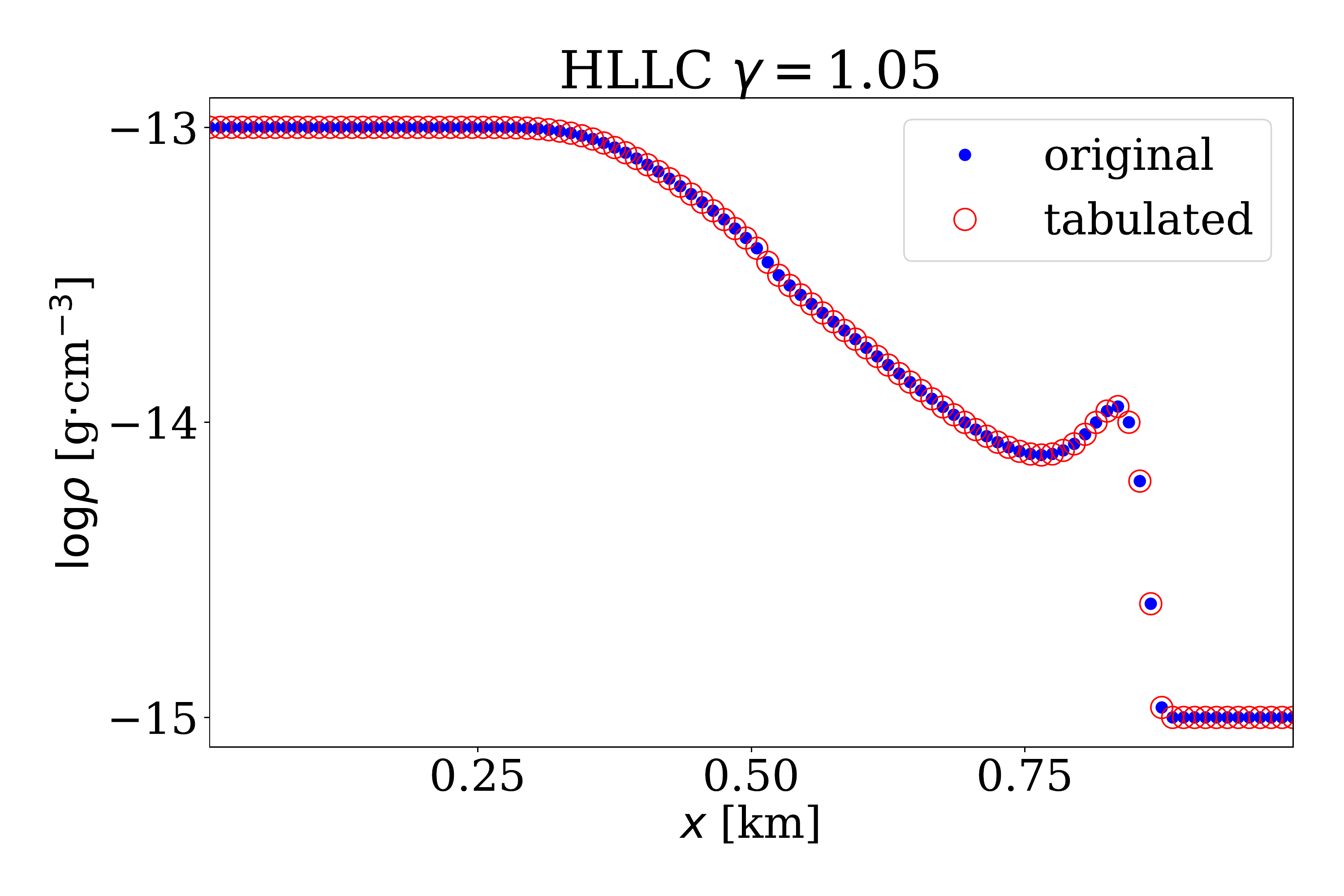}
    \includegraphics[width=0.495\textwidth]{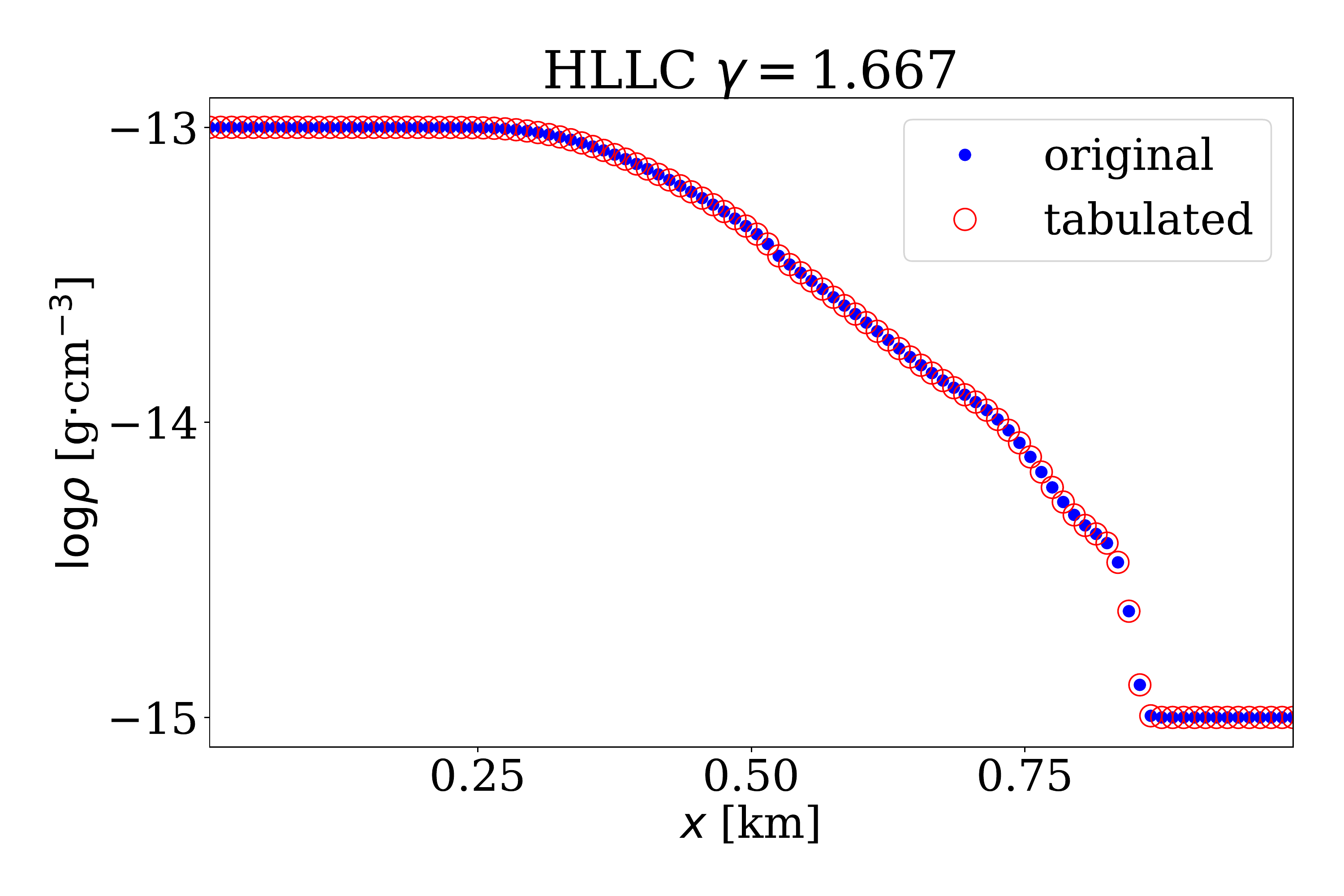}
    \caption{Solutions of the shock-tube test 2 in Table \ref{tab:hllc} of the HLLC general EoS Riemann solver and the original HLLC Riemann solver  at $t=0.012$s. EoS tables of perfect gas have been provided to the HLLC general EoS Riemann solver.}
    \label{fig:hllcvshllc1}
\end{figure}

For the same reason, as stated in Section \ref{sec:exactvsexact}. The HLLC general EoS Riemann solver should reduce to the original HLLC Riemann solver when the EoS tables of a perfect gas are used. The EoS is solved with the interpolation method described in Section \ref{sec:hllceossolver}. We carry out two tests that have the solution structure of a shock and rarefaction waves with $\gamma=1.05$ and $1.667$. The left and right states are listed in Table \ref{tab:hllc}. The coordinate that separates the initial left and right states is $x=0.5$ km, and the simulation domain is $[0,1]$ km. We use the first order Godunov scheme \cite{godunov1959} and let $\text{CFL}=0.7$. There are 100 cells in our simulation and Figure \ref{fig:hllcvshllc1}.
From Figure \ref{fig:hllcvshllc1}, we can tell that the solutions from the two HLLC Riemann solvers also produce very similar results, confirming that the HLLC general EoS Riemann solver reduces to the original HLLC Riemann solver for a perfect gas.

\begin{table}[!h]
    \centering
    \begin{tabular}{|c|c|c|c|c|c|c|c|}\hline
      ID  & $t_{\rm{max}}$ & $\rho_{\rm L}$  &   $v_{\rm L}$  & $T_{\rm L}$  & $\rho_{\rm R}$  & $v_{\rm R}$  &  $T_{\rm R}$ \\
          &  [s] & [\gcmc]  &   [\kms]  &   [K]  &   [\gcmc]  &  [\kms]  &   [K]  \\\hline
      2   & 0.012 &$10^{-13}$   &    0   &    $20000$  &  $10^{-15}$  &  0   &  $300$  \\\hline
    \end{tabular}
    \caption{Simulation run-times ($t_{\rm{max}}$) and initial left and right states of the Riemann problem tests. $\mu=1.3$.}
    \label{tab:hllc}
\end{table}

\subsection{Comparison of the exact solution to solutions of the HLLC general EoS Riemann solver with interpolation EoS solver and with analytic EoS solver}\label{sec:hllcvsexact}

In this section, we use pure hydrogen gas with two binding energies (\ch{H2} disassociation and \ch{H} ionization) as an example of realistic gas. Therefore, four species are considered (\ch{H2}, \ch{H}, \ch{H^+}, and \ch{e^-}, see Appendix \ref{sec:eos} for more information). We compare three kinds of numerical solvers (with different EoS schemes) on four tests using our pure hydrogen gas EoS: the exact solution calculated by the algorithm in Section \ref{sec:psolver}, the HLLC general EoS Riemann solver with an interpolated EoS (discussed in Section \ref{sec:hllceossolver}), and the HLLC general EoS Riemann solver with an analytic EoS solver. The analytic EoS solver solves the EoS by calculating thermodynamic variables at run-time and is explicitly described in Appendix \ref{sec:eos}. We show that the solution of the HLLC general EoS Riemann solver with the two EoS solvers approaches the solution of the exact solution as the $L^{1}$ norm of their difference decreases monotonically with the increasing resolution. Explicitly, we calculate,
\begin{equation}
    \delta_{N}(f)\equiv \frac{L}{N} \sum_{i=1}^{N} \left|f_{N}(x_i)-f_{\rm exact}(x_i)\right|
\end{equation}
where $f_{N}$ is the solution of any specific field, $N$ is the number of points. $f_{\rm exact}$ is the solution of exact general EoS Riemann solver of any specific field, and $L=1$ km is the length of the simulation domain. In this research, we choose the number of cells $N=100,200,400,800$ (we shall use the term ``resolution" interchangeably) and the density field to calculate the corresponding $L^{1}$ norm. We designed four tests to demonstrate that the HLLC general EoS Riemann solver can handle the 1D hydrodynamic problem with multiple phase transitions well, which are listed in Table \ref{tab:hllcvsexact}. The coordinate that separates the initial left and right states for all these four tests is $x=0.5$ km and the simulation domain is $[0,1]$ km. We use the first order Godunov scheme and let $\text{CFL}=0.7$.

Test 2 is a strong shock tube test (same as the initial condition in Table \ref{tab:hllc}). Its left state consists of fully ionized hydrogen, and its right state consists of purely molecular hydrogen. Test 3 is an asymmetric two-shock-waves test. Its left and right gases are all in a purely molecular state, but the shocked gas would be (partially) ionized. Test 4 is a symmetric two-rarefaction-wave test. Its left and right states are all fully ionized. The rarefied gas in the middle region will recombine and form atomic hydrogen. Test 5 is also an asymmetric two-shock-wave test. Its difference from test 3 is that the shocked gas will be more ionized.

\begin{table}
    \centering
    \begin{tabular}{|c|c|c|c|c|c|c|c|}\hline
      ID  & $t_{\rm{max}}$  & $\rho_{\rm L}$      & $v_{\rm L}$ & $T_{\rm L}$ & $\rho_{\rm R}$    & $v_{\rm R}$ &  $T_{\rm R}$ \\
          & [s]   &  [\gcmc]            &   [\kms]    &   [K]       &   [\gcmc]         &  [\kms]     &   [K]  \\\hline
      2   & 0.012 &$10^{-13}$   &    0   &    $20000$  &  $10^{-15}$  &  0   &  $300$  \\\hline
      3   & 0.100 & $2\times10^{-15}$   &  40         &      $800$  & $10^{-15}$        &  -60        &  $800$  \\\hline
      4   & 0.008 & $2\times10^{-11}$   & -30         &    $15000$  & $2\times10^{-11}$ &  30         &  $15000$  \\\hline
      5   & 0.070 & $1.2\times10^{-15}$ &  55         &      $500$  & $10^{-15}$        &  -65        &  $500$  \\\hline
    \end{tabular}
    \caption{Run-times ($t_{\rm max}$) and initial left and right states of the four tests with pure hydrogen gas. The tests evolve a strong-shock, asymmetric two-shock, symmetric two-rarefaction, and asymmetric two-shock-wave respectively. More detailed descriptions are in the text.}
    \label{tab:hllcvsexact}
\end{table}

When using Riemann solvers with Godunov scheme, we can get the solution of $\rho$, $v$, and $\epsilon$ after each step. The tabulated EoS solver calculates $p,\ T,\ \Gamma$, and $\mu$ with the bi-linear interpolation method discussed in Section \ref{sec:hllceossolver}. With $\mu$ and $\rho$, $n_{\ch{H2}},\ n_{\ch{H},\ n_{\ch{H+}}}$, and $n_{\ch{e-}}$ can be calculated analytically. We also calculate
\begin{align}
    \chi_{\ch{H2}}&=\frac{2n_{\ch{H2}}}{2n_{\ch{H2}}+n_{\ch{H}}+n_{\ch{H+}}}  \label{eqn:concentration1}\\
    \chi_{\ch{H}}&=\frac{n_{\ch{H}}}{2n_{\ch{H2}}+n_{\ch{H}}+n_{\ch{H+}}}  \label{eqn:concentration2}\\    \chi_{\ch{H+}}&=\frac{n_{\ch{H+}}}{2n_{\ch{H2}}+n_{\ch{H}}+n_{\ch{H+}}} \label{eqn:concentration3}
\end{align}
as the concentration percentage of each hydrogen species. The analytic EoS solver first calculates $T$ by solving the inverse problem of $T=T(\rho,\epsilon)$, then it calculate $p,\ \Gamma$ and all species number density $n_{i}$ by solving the forward problem of $z=z(\rho,T)$ where $z$ should be replaced by the interested variable. $\mu$ is calculated by $\mu=\rho/(m_{\mu}\sum_{i}n_{i})$ where $m_{\mu}$ is the atomic mass unit.

\begin{figure}
	\centering
    \includegraphics[width=\textwidth]{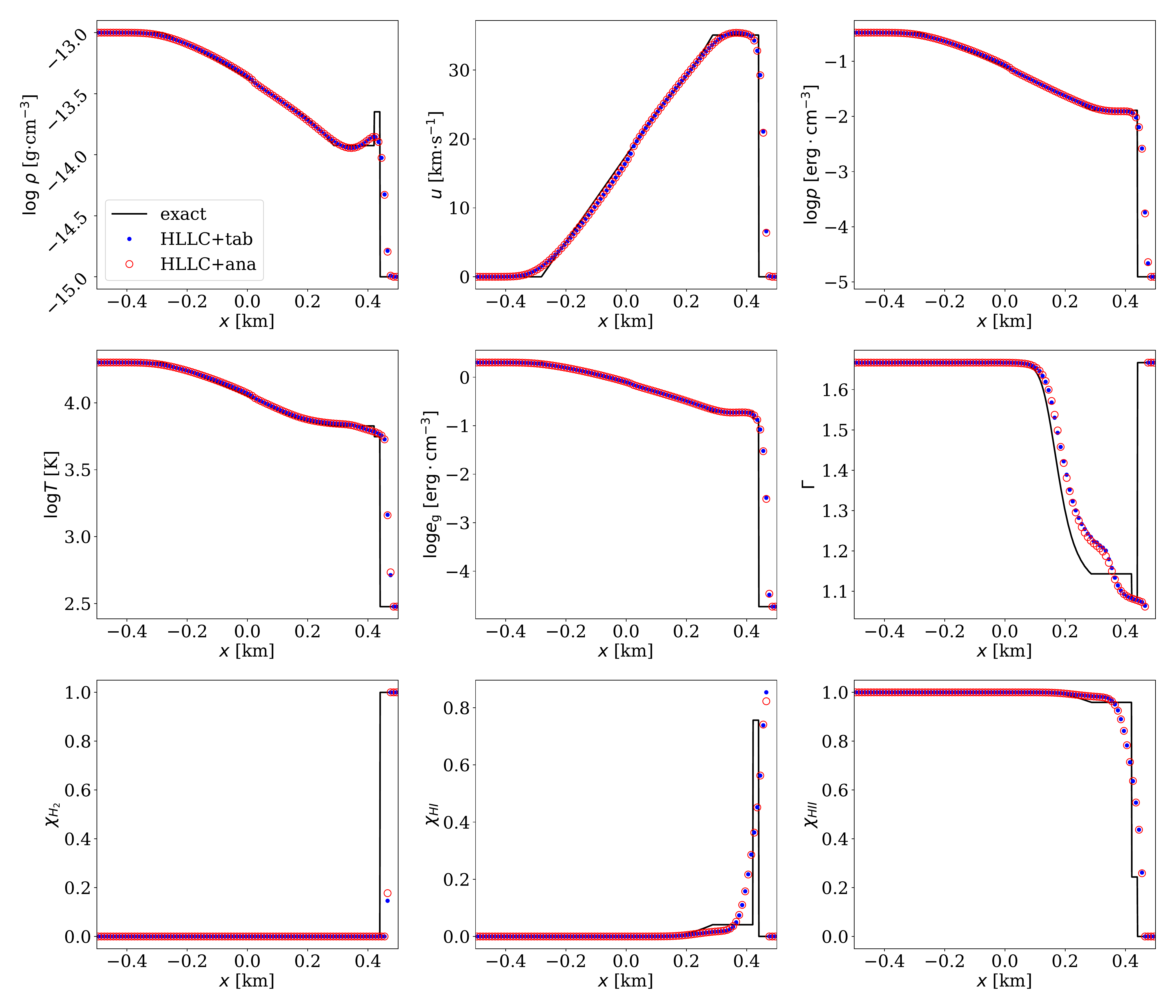}
    \caption{The solutions of test 2 (see Table~\ref{tab:hllcvsexact}). The results from the exact general EoS Riemann solver (black), HLLC general EoS Riemann solver with tabulated EoS (blue) and analytic EoS (red), $N=100$. The first row from the left to the right are the density $\rho$ in \gcmc, the velocity $v$ in \kms, and the pressure $p$ in bar. The panels in the second row from the left to the right are the temperature $T$ in K, the internal energy per volume $e_{\rm g}=\rho\epsilon$ in \ergcm, and the adiabatic index $\Gamma$. The panels in the third row from the left to the right are the concentration of hydrogen in \ch{H2}, \ch{H}, and \ch{H+}, respectively. The black, blue, and red dots represent the solution of the exact general EoS Riemann solver, HLLC general EoS solver with tabulated EoS solver and analytic EoS solver, respectively.} 
    \label{fig:eossolver1}
\end{figure}

\begin{figure}
	\centering
    \includegraphics[width=\textwidth]{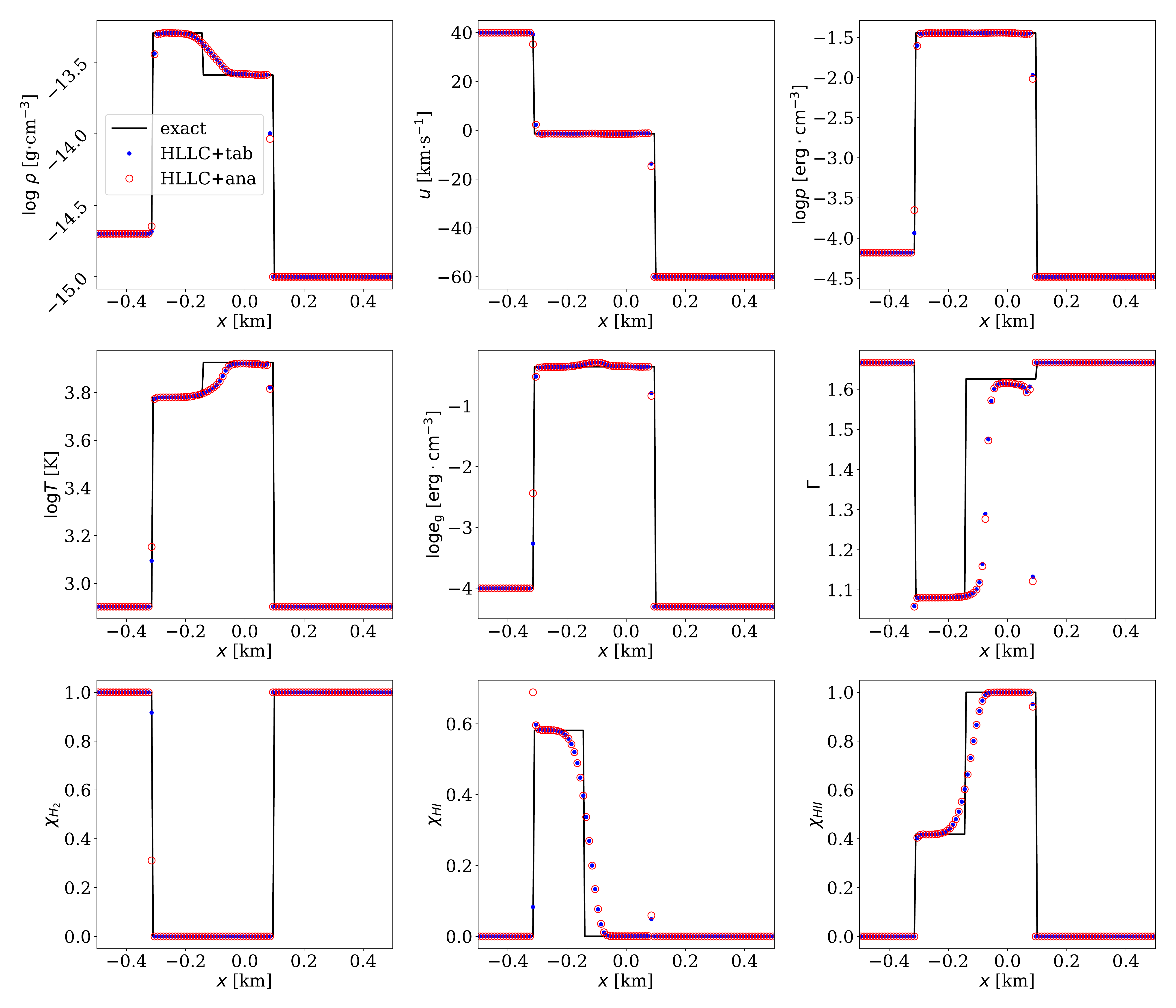}
    \caption{The solutions of test 3 (see Table~\ref{tab:hllcvsexact}). Same layout as Figure \ref{fig:eossolver1}.}
    \label{fig:eossolver2}
\end{figure}

\begin{figure}
	\centering
    \includegraphics[width=\textwidth]{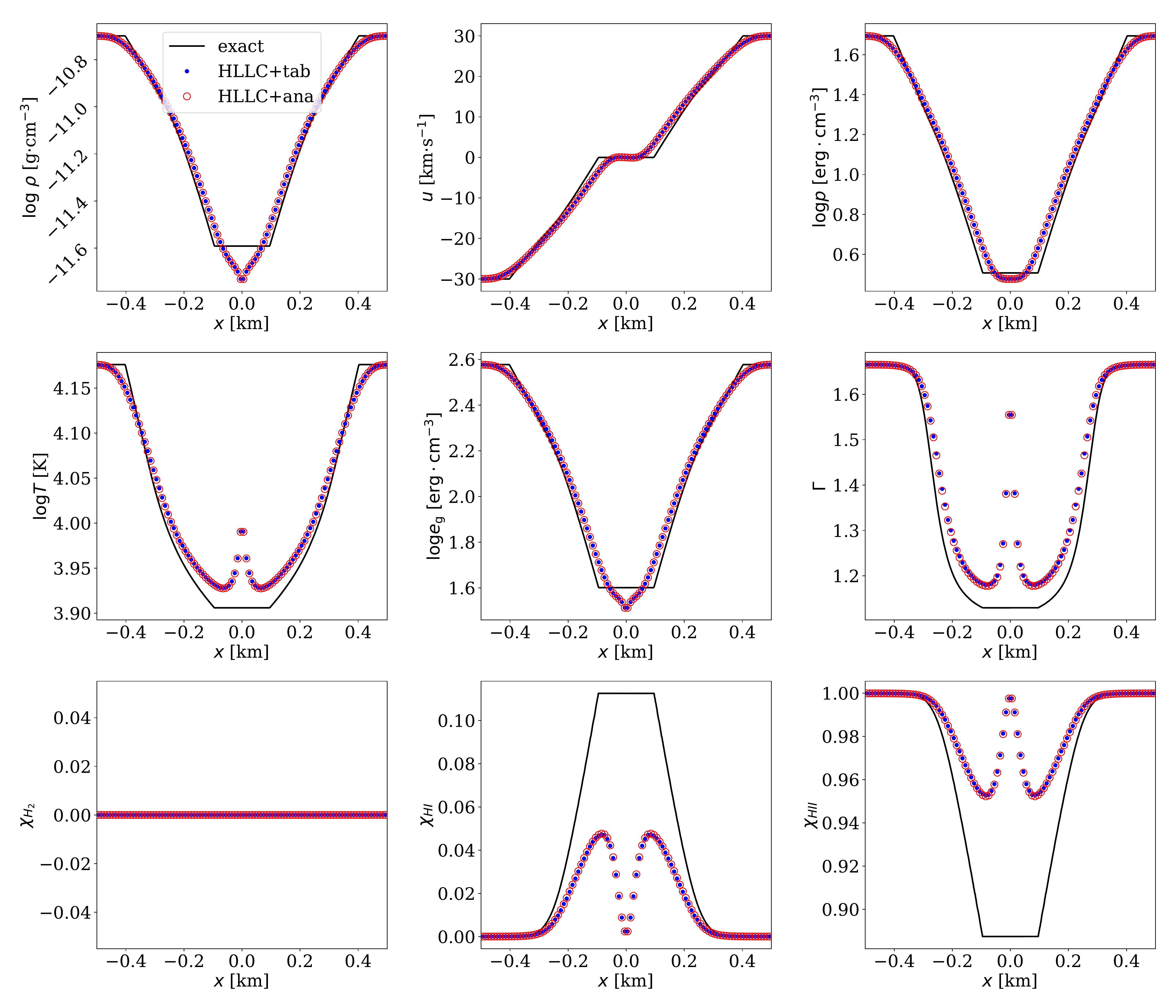}
    \caption{The solutions of test 4 (see Table~\ref{tab:hllcvsexact}). Same layout as Figure \ref{fig:eossolver1}.}
    \label{fig:eossolver3}
\end{figure}

\begin{figure}
	\centering
    \includegraphics[width=\textwidth]{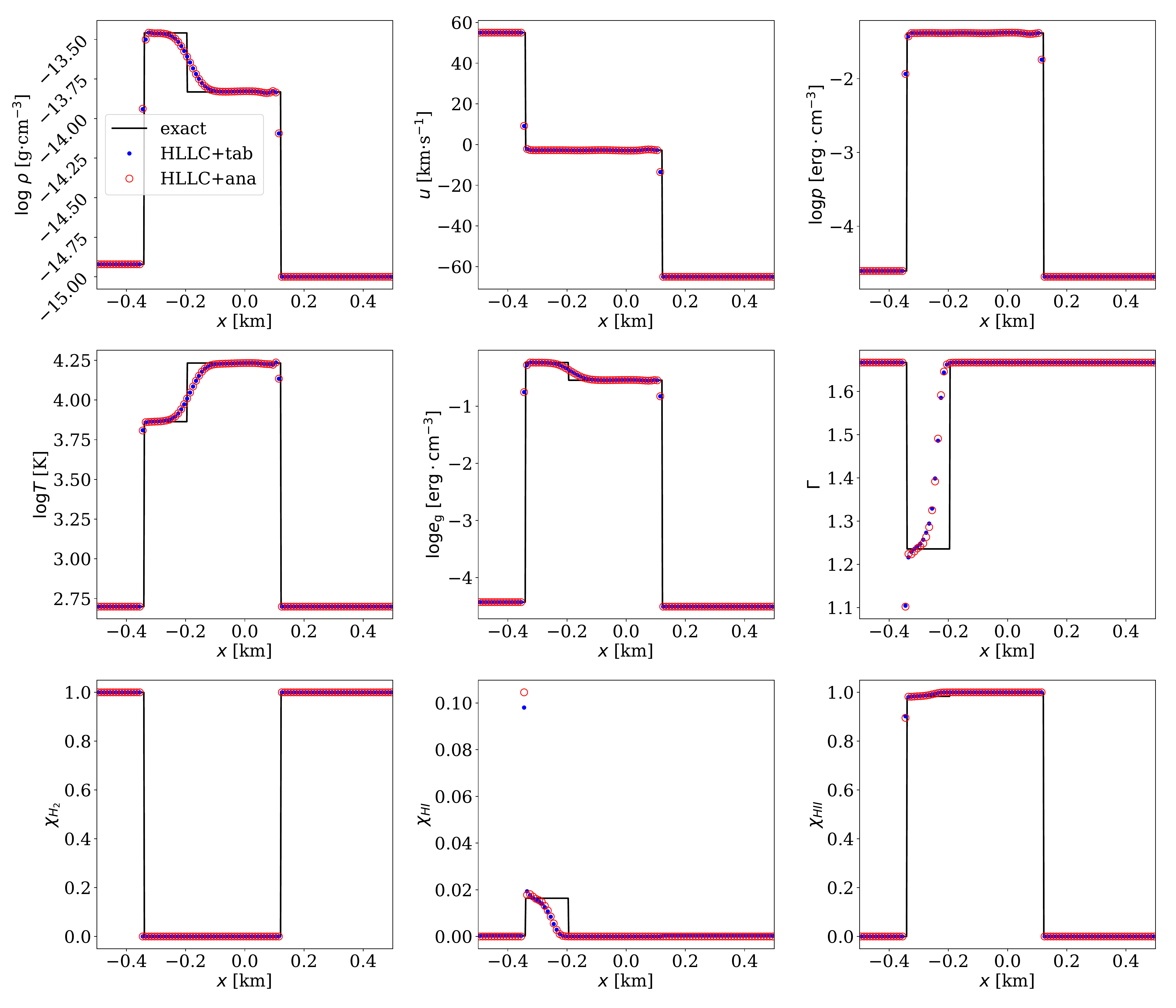}
    \caption{The solutions of test 5 (see Table~\ref{tab:hllcvsexact}). Same layout as Figure \ref{fig:eossolver1}.}
    \label{fig:eossolver4}
\end{figure}

The solutions produced by the three aforementioned Riemann solvers with $N=100$ for these four tests are shown in Figure \ref{fig:eossolver1}, \ref{fig:eossolver2}, \ref{fig:eossolver3}, and \ref{fig:eossolver4}. The arrangement of the panels in all four figures are the same. The solutions of the tabulated EoS solver and the analytic EoS solver have little difference.

In Figure \ref{fig:eossolver1}, the shocked gas is compressed, and the solution structure resembles a low $\gamma$ perfect gas. Since the thickness of the compressed gas is small, at $N=100$, there are only two cells inside the compressed region. The solution of the HLLC solvers gives much ``diffused" density profiles. Nonetheless, the position of the shock is well captured, and the solution of $u,\ p,\ T$, and $e_{\rm g}$ are similar to the exact solution. $\Gamma$ is smooth but different from the exact solution. We think the $\Gamma(\log\rho,\log p)$ table is well resolved as $d\log p=0.011$ which correspond to $2.56\%$ change in pressure. This point is also pronounced in the analytic EoS solver as the convergence criterion requires that the change of $T$ should be less than $0.001\%$, or $dT<1$ K at $T=10000$ K. Therefore, the EoS solvers are less likely to be the cause of this defect. There could be some fundamental limitation in the current HLLC algorithm or improvement of the HLLC Riemann solver may be made in the future to improve the accuracy of the solution. $\chi_{\ch{H2}},\ \chi_{\ch{H}}$, and $\chi_{\ch{H+}}$ are related to the calculation of $\mu$ in the tabulated EoS solver or $T$ in the analytic EoS solver. Both approximate solutions roughly caption the position of the ionization front. The sensitive relation of number density to $\mu$ or $T$ and the diffusive nature of the Godunov scheme lead to the difference between the approximate solutions and the exact solution.

In Figure \ref{fig:eossolver2}, the position of the middle wave in the approximate solution of $\rho$ and $T$ does not match the exact solution very well. There is a little bump in the solution of $e_{\rm g}$ while $p$ does not have. This situation cannot happen in perfect gas since $e_{\rm g}$ and $p$ has a linear relation in perfect gas. The bump in $e_{\rm g}$ could be related to some defect in the HLLC solver. $\Gamma$ does not give accurate solution and the reason could be the same as test 2. The solutions of $\chi_{\ch{H2}},\ \chi_{\ch{H}}$, and $\chi_{\ch{H+}}$ are more satisfactory.

What we see in Figure \ref{fig:eossolver3} represents the defect of Godunov scheme. The ``heating" at the center of the symmetric rarefaction waves and the ``dip" in density are typically found in Godunov scheme (see e.g. \cite{ryu1995} for the magneto-hydrodynamic case and Section 6.4.2, Chapter 6 of \cite{toro2013}).

In Figure \ref{fig:eossolver4}, $\rho,\ u,\ p,\ T$, and $e_{\rm g}$ are all very satisfactory. Differences can be found in the solutions of $\chi_{\ch{H}}$ between the exact and approximate solutions. This is because the gas is on the edge of becoming fully ionized therefore $\chi_{\ch{H}}$ is very sensitive to $\rho$ and $T$.

An interesting thing to note is that both test 3 and test 5  (Figures \ref{fig:eossolver2} and \ref{fig:eossolver4} respectively) consist of two strong shocks. The ``accuracy" of approximate Riemann solvers seem very different. This implies that the HLLC general EoS Riemann solvers discussed in this paper may not be very accurate when dealing with partially ionized or disassociated gas, or gas that is right in phase transitions. One reason may be that we are not finding an optimal $p_{*}$ with the adaptive non-iterative Riemann solver. However, using the exact general EoS Riemann solver to calculate $p_{*}$ would be too expensive. Another possibility is that the change of $\epsilon$ during transition is too stiff. 
Alternatively, this error could be cause by the fact that the discontinuities spend most of their time in between cell interfaces. When this occurs the cells containing the discontinuity will contain both neutral and ionized gas and be forced to assume some intermediate phase introducing errors.
Nonetheless, the HLLC general EoS Riemann solvers proposed in this paper are still available options in general EoS hydrodynamics, and these test are designed to probe the limits of our code.

We also study the efficiency of the HLLC general EoS Riemann solver with the analytic EoS solver and tabulated EoS solver. We compare them to the original HLLC Riemann solver at the same resolution of $N=800$. We calculate the solution of the nine quantities shown in Figure \ref{fig:eossolver1}-\ref{fig:eossolver4} with the general EoS Riemann solvers and only calculate the solution of $\rho,\ u,\ p,\ T$, and $e_{\rm g}$ with the original HLLC Riemann solver. The CPU time for the four tests in Table \ref{tab:hllcvsexact} are listed in Table \ref{tab:time}.

\begin{table}
    \centering
    \begin{tabular}{|c|c|c|c|c|}\hline
         &  test2  &  test3   &  test4   &  test5  \\\hline
      $t_{\rm{o}}$ [s]  &  $2.250\times10^{-1}$  &  $2.546\times10^{0}$  &  $1.750\times10^{-1}$  &  $2.186\times10^{0}$  \\\hline
      $t_{\rm{i}}$ [s]   &   $1.729\times10^{1}$  &  $8.849\times10^{1}$  &  $2.318\times10^{1}$  &   $6.175\times10^{1}$    \\
       speed   &  76.84  &  34.76  &  132.5  &  28.25   \\\hline
      $t_{\rm{a}}$ [s]   &   $5.472\times10^{0}$   &  $5.335\times10^{1}$  &  $1.022\times10^{1}$   &  $2.846\times10^{1}$  \\
      speed  &  24.32  &  20.95  &  58.40  &   13.01    \\\hline
    \end{tabular}
    \caption{CPU time of the three HLLC Riemann solvers of test 2-5 in Table \ref{tab:hllcvsexact} in second. We show the CPU time of the original HLLC Riemann solvers, the HLLC general EoS Riemann solver with interpolation method, and the HLLC general EoS Riemann solver with analytic EoS solver. The ratio of the CPU time of the two HLLC general EoS Riemann solvers to the original HLLC Riemann solver are shown below each solver.}
    \label{tab:time}
\end{table}

The HLLC general EoS Riemann solver with analytic EoS solver and tabulated EoS solver are about 13-58 times slower and 35-132 times slower than the original HLLC Riemann solver, respectively. Since test 2 has one rarefaction wave and test 4 has two rarefaction waves, Table \ref{tab:time} tells us that the HLLC general EoS Riemann solver is less efficient in calculating rarefaction waves. We further conclude some advantages and disadvantages of each general EoS Riemann solvers below:
\begin{enumerate}
    \item If the span of the EoS domain is big in $(\log\rho,\log p)$, the tabulated EoS Riemann solver becomes less accurate or slower. This is because the table needs to be enlarged to maintain the same resolution (or accuracy) and can slower down the interpolation process significantly which is mainly due to the increase of the memory accessing time.
    \item If the gas consists of \ch{H2}, \ch{H}, \ch{H+}, \ch{He}, \ch{He+}, \ch{He^{2+}}, and \ch{e-}, the analytic EoS solver needs to solve two equilibrium system and each one of them may contain a quartic equation. This can slow down the analytic EoS solver significantly.
    \item If the EoS is hard to be approximated by the analytic EoS solver, for example, the rotational and vibrational degrees of freedom in \ch{H2}, and the problem is sensitive to these two degrees of freedom, the tabulated EoS solver becomes only option.
    \item When using the tabulated EoS solver, one has to estimated the range of $\log\rho$ and $\log p$ and choose the suitable EoS tables before solving the problem. An over-estimation could reduce the efficiency and an under-estimation may lead to unexpected outcome. The analytic EoS solver does not have this limitation.
\end{enumerate}
With the above conclusions, we propose that a combination of the analytic EoS and tabulated EoS solvers may be more favorable. We have not implemented other high order interpolation methods in the tabulated EoS solver, but we think it very much worth investigating as long as the interpolation method preserves the monotonicity of the tables.

\begin{figure}
    \centering
    \includegraphics[width=0.8\textwidth]{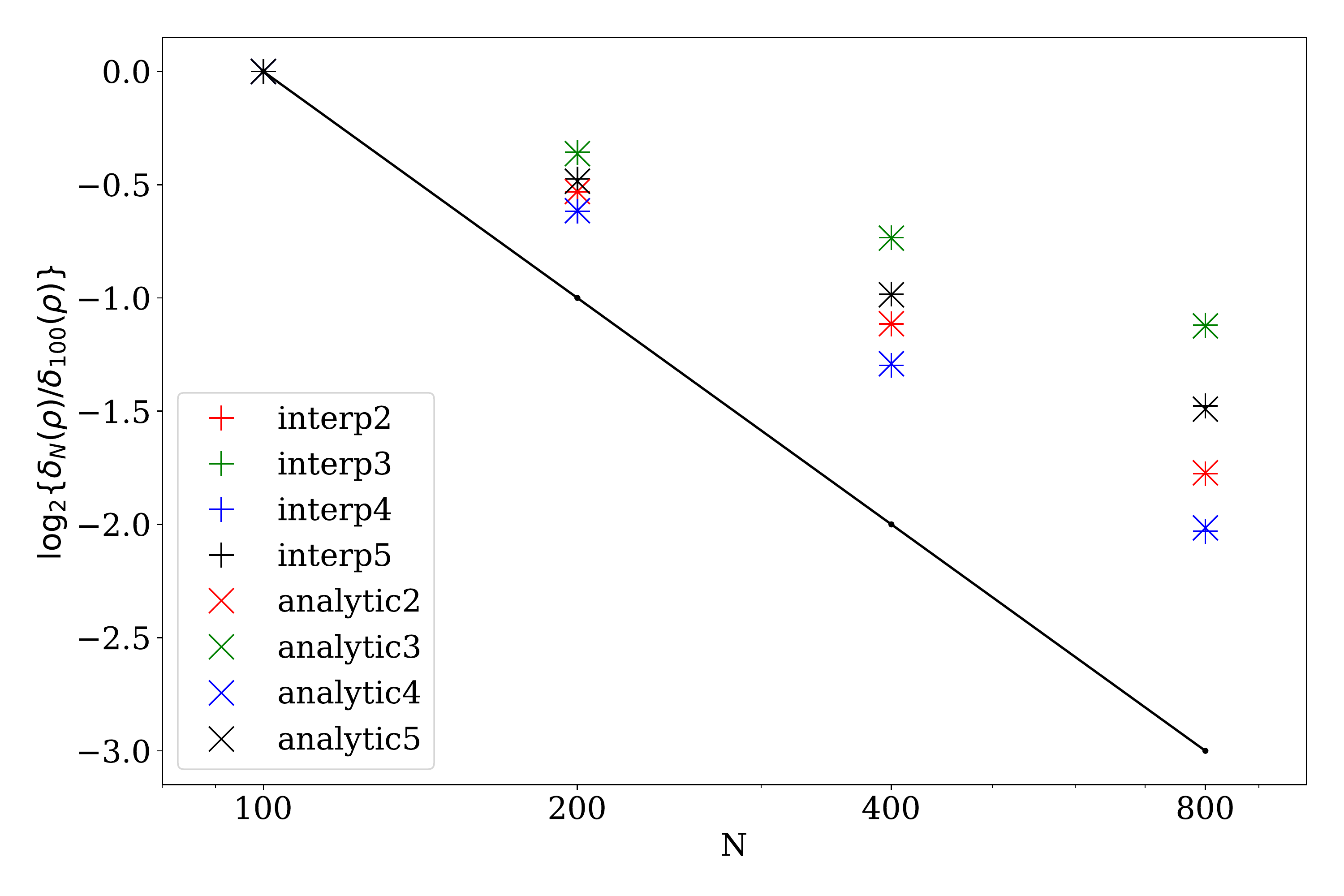}
    \caption{$\log_{2}\{\delta_{N}(\rho)/\delta_{100}(\rho)\}$ v.s. $N$. Different colors represent the value of the corresponding tests listed in Table \ref{tab:hllcvsexact}.}
    \label{fig:convergence}
\end{figure}

At the end of this section, we plot the normalized value $\log_{2}\{\delta_{N}(\rho)/\delta_{100}(\rho)\}$ of each test for both HLLC general EoS Riemann solver with tabulated and analytic EoS solvers in Figure \ref{fig:convergence}. The black line represents the linear convergence rate, which is the highest convergence rate one can achieve because the exact solutions have discontinuities (except for the two rarefaction waves). Both solvers show a clear trend of decreasing $\log_{2}\{\delta_{N}(\rho)/\delta_{100}(\rho)\}$ with increasing resolution $N$ but their convergence rate is less than the linear convergence rate. The worst convergence rate happens in test 3 that has two strong shocks with partially ionized middle states.

\subsection{A 2D simulation}

\begin{figure}
    \centering
    \includegraphics[width=\columnwidth]{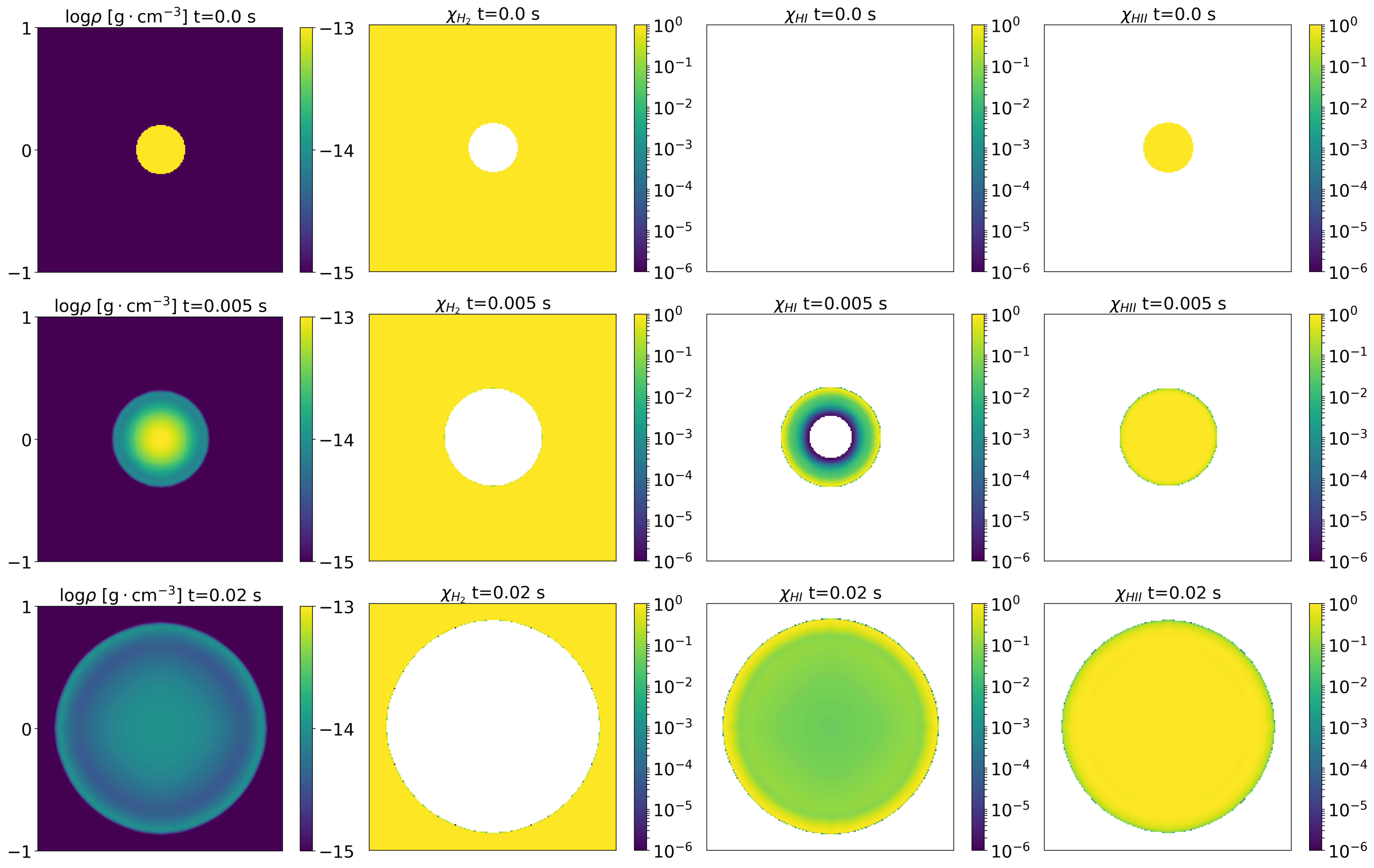}
    \caption{Shock test in 2D using the HLLC general EoS Riemann solver with analytic EoS solver.}
    \label{fig:2deos}
\end{figure}

Figure \ref{fig:2deos} shows a shock test in 2D. We use the HLLC general EoS Riemann solver with analytic EoS solver. The simulation domain is $[-1,1]\times[-1,1]$ km$^{2}$ with a fixed resolution of $200\times200$. We place a circle with $\rho=10^{-13}$ \gcmc and $T=20000$ K at the origin and its radius is $0.2$ km at $t=0$ s. The ambient has an initial condition of $\rho=10^{-15}$ \gcmc and $T=300$ K. In this 2D simulation, CFL$=0.7$ and we confirm that $(v_{x}/dx+v_{y}/dy)\Delta t<1$ at all time in the domain.  $v_{x}$ and $v_{y}$ in the x and y direction. $dx=dy=0.01$ km is the grid size and $\Delta t$ is the time step. The 2D integration method used in this simulation is symmetric Strang splitting \citep{Strang1968}.

Each row in Figure \ref{fig:2deos} shows $\rho,\ \chi_{\ch{H2}},\ \chi_{\ch{H},}$ and, $\chi_{\ch{H+}}$ at different time. $\chi=0$ is represented by the white area as we divide the EoS into six regions and set the number density of some species to be $0$ in some region. This simulation shows an outward shock and a inward rarefaction waves. The center of the ``fireball" cools down and $\ch{H}$ appears. The shocked gas is highly compressed due to the geometric effect and the higher compressibility of realistic gas.

\section{Conclusions and discussions}\label{sec:con}

Following the same principle of the original HLLC Riemann solver, we devise an HLLC general EoS Riemann solver that can solve realistic gas dynamics with drastically varying $C_{V}$ and $C_{P}$. We intentionally keep the modification of the original HLLC Riemann solver to a low extent to achieve a fast speed.

To examine the Riemann problem for a realistic fluid and determine the accuracy of the HLLC general EoS Riemann solver, we implemented an exact general EoS Riemann solver based on \cite{colella1985}. We improved the robustness of the root-finding algorithm. The existence and uniqueness of the solution are guaranteed if the EoS tables and the interpolated ones satisfy the two monotone condition in Equation \ref{eqn:unique1}-\ref{eqn:unique2}. The new exact general EoS Riemann solver can be used even when a convex condition $\left(\frac{\partial^{2}p}{\partial\rho^{2}}\right)_{s}>0$ is not satisfied. We applied the exact general EoS Riemann solver to the Godunov scheme in Section \ref{sec:exactgodunov}.
To examine the correctness of our new algorithm, we compared our exact general EoS Riemann solver to the exact ideal gas Riemann solver for shock tube tests of perfect gases with $\gamma=\Gamma=1.05$ and $\gamma=\Gamma=1.667$. 
The solutions are comparable, confirming that our exact general EoS Riemann reduces to the original exact ideal gas Riemann solver when using the corresponding perfect gas EoS.
We also compared our HLLC general EoS Riemann solver to the original HLLC Riemann solver by running tests with perfect gas EoS, resulting in comparable solutions. Therefore, our HLLC general EoS Riemann solver reduces to the original HLLC Riemann solver in the case of a perfect gas.

In Section \ref{sec:hllcvsexact}, we compare the solution of the HLLC general EoS Riemann solver with two different EoS solvers to the exact solution calculated by the algorithm in Section \ref{sec:psolver}. One EoS solver is called the tabulated EoS solver; it calculates the thermodynamic variables by interpolating 2D tables with a bi-linear interpolation method (Section \ref{sec:hllceossolver}). Another EoS solver is called the analytic EoS solver; it calculates the thermodynamic variables by solving EoS at run-time (see Appendix~\ref{sec:eos} for details).

We find reasonably consistent results between the approximate and exact Riemann solvers for the shock-tube and two-shock tests. The HLLC solver can resolve the contact discontinuity, which is different from the HLL solver \cite{kannan2019}. The location and the strength of the shock can be well captured. Although discrepancy is found in the two-rarefaction-wave test, we think defect is rooted in the Godunov scheme. By comparing two tests (test 3 and 5) that have two strong shock waves with different ionization level, we infer that the HLLC general EoS Riemann solver still has difficulty in resolving shocked gases that is in the middle of a phase transition. The efficiency of the new HLLC general EoS Riemann solver with both EoS solvers are compared to the original HLLC Riemann solver. We find that the analytic EoS solver is roughly 20 times slower than the original HLLC solver, while the tabulated EoS solver is roughly 60 times slower than the original HLLC solver. However, we point out that these multiples are not constant depending on the problem being solved in Section \ref{sec:hllcvsexact}. In short, tabulated EoS solver becomes slower in problems with a large range of $\log\rho$ or $\log p$, and analytic EoS solver becomes slower in problems with many species in equilibrium. A combination of the two EoS solvers may become favorable in problems that have a large range of $\log\rho$, $\log p$, and many species present at the same time. In multi-physics simulations that include self-gravity (Poisson equation), complicated radiation transfer \cite{davis2012,jiang2012}, etc., hydro-step only takes a fraction of the total computational time. If the temperature is critical in those simulations, the HLLC solvers proposed in this research are still useful even though they are much more expensive than the perfect gas HLLC Riemann solver. Convergence study has been performed by comparing the solutions of our HLLC general EoS Riemann solver to the exact solution. We use the $L^{1}$ norm of density to demonstrate that the HLLC Riemann solver approaches the exact solution with increasing resolution. We found that the convergence rates are below the linear convergence rate, with the worst one happens in the case of two shocks with partially ionized gas.

\section*{Acknowledgments}

We thank James Stone, Kengo Tomida, Yan-Fei Jiang, Jonathan Carroll-Nellenback, Natasha Ivanova, and Xuening Bai for their useful discussions and insight generated from their work. We also thank the anonymous referee who's feedback lead to significant improvements of this paper.
ZC is grateful to the Horton Fellowship at the University of Rochester and the CITA National Postdoctoral Fellowship for financial support.
MC gratefully acknowledges support from the Institute for Advanced
Study, NSF via grant AST-1515763, and NASA via grant
14-ATP14-0059.
AF appreciates financial support from the US Department of Energy through grant DE-SC0001063 ,the NSF through grant AST-1515648, and the Space Telescope Science Institute  through grant HST-AR-12832.01-A.

\begin{appendices}

\section{Basic thermodynamics: differentiating $\gamma$ and $\Gamma$}\label{sec:thermaldynamics}

We use $\gamma$ to represent the ratio of specific heats, i.e. $\gamma=C_{p}/C_{V}$ where $C_{p}$ and $C_{V}$ are the constant pressure heat capacity and the constant volume heat capacity, respectively. We use $\Gamma$ to represent the adiabatic index which is defined as
\begin{equation}
    \Gamma=\left(\frac{\partial\ln p}{\partial\ln\rho}\right)_s
\end{equation}
and the adiabatic sound speed
\begin{equation}
    a=\sqrt{\left(\frac{\partial p}{\partial\rho}\right)_s}=\sqrt{\frac{\Gamma p}{\rho}},
\end{equation}
where the subscript $s$ means constant specific entropy.

In this section, we will show that $\gamma=\Gamma$ for perfect gas but not necessarily for realistic gas.

First, we shall derive $\gamma$. A general thermal dynamic relation that consider multiple species is:
\begin{align}
    dU&=Tds-pdV+\sum_{i}\mu_{i}dN_{i}, \\
    dH&=Tds+Vdp+\sum_{i}\mu_{i}dN_{i},
\end{align}
where $i$ stands for different species. $U,\ H,\ s,\ \mu_{i},\ $and $N_{i}$ ($\mu_{i}$ is not to be confused with the mean atomic weight $\mu$) are the internal energy, enthalpy, entropy, chemical potential, and particle number, respectively. By the definition of $C_{V}$ and $C_{p}$,
\begin{align}
    C_{V}&=\left(\frac{dU}{dT}\right)_{s}, \label{eqn:CV}  \\
    C_{p}&=\left(\frac{dH}{dT}\right)_{s}=\left(\frac{d(U+pV)}{dT}\right)_{s,p},
\end{align}
therefore,
\begin{equation}\label{eqn:gammacompare1}
    \gamma=\frac{C_{p}}{C_{V}}=\left(\frac{d(U+pV)}{dU}\right)_{s,p}.
\end{equation}
If we further consider ideal gas with fixed species number, i.e. $dN_{i}=0$, then $pV=Nk_{b}T$ where $N$ is the total particle number. The key step in deriving $\Gamma$ is to write $dT$ in terms of $dp$ and $d\rho$ in adiabatic process,
\begin{align}
    dT&=\frac{Vdp}{Nk_{b}}+\frac{pdV}{Nk_{b}}  \\\label{eqn:dT}
    &=\frac{dp}{\rho Nk_{b}}-\frac{pd\rho}{\rho^{2}Nk_{b}},
\end{align}
where we have used $V=1/\rho$. Substitute Equation \ref{eqn:dT} into Equation \ref{eqn:CV} and rearrange the equation,
\begin{equation}
    \frac{dp}{p}=\frac{C_{V}+Nk_{b}}{C_{V}}\frac{d\rho}{\rho},
\end{equation}
therefore,
\begin{equation}\label{eqn:gammacompare2}
    \left(\frac{dp}{d\rho}\right)_{s}=\frac{C_{V}+Nk_{b}}{C_{V}}\frac{p}{\rho}.
\end{equation}
Compare Equation \ref{eqn:gammacompare1} to Equation \ref{eqn:gammacompare2}, we can conclude that, for constant species number ideal gas (and perfect gas),
\begin{equation}
    \left(\frac{dp}{d\rho}\right)_{s}=\frac{\gamma p}{\rho},
\end{equation}
therefore $\gamma=\Gamma$, which is a coincidence. For realistic gas, $dN_{i}\neq0$ thus even with ideal gas assumption,
\begin{equation}
    dT=\frac{dp}{\rho (\sum_{i}N_{i})k_{b}}-\frac{pd\rho}{\rho^{2}(\sum_{i}N_{i})k_{b}}-\sum_{i}\frac{pdN_{i}}{\rho(\sum_{i}N_{i})^{2}k_{b}}
\end{equation}
becomes nontrivial and it is even more difficult to express $dN_{i}$ in terms of $dp$ and $d\rho$ since $dN_{i}$ depends on $(\rho,T)$ in general and is solved with Saha equation. The strategy we used to derive $\Gamma$ fails. In fact, $\gamma=\Gamma$ may not hold anywhere in $(\rho,T)$ for realistic gas. We will show the calculation of $\Gamma$ for a simplified pure hydrogen gas in Appendix \ref{sec:eos}.

\section{Quasilinear Approximation}\label{sec:qlin}
The computation of the characteristics (i.e. speeds in the Riemann fan) is typically done with the quasilinear approximation. 

\subsection{Conservative Form}\label{sec:conserveform}
First we start with the Euler equations in differential, conservative form:
\begin{subequations}
\begin{align}
\pder{\rho}{t}+\pder{}{x}(\rho u)&=0\\
\pder{}{t}(\rho u)+\pder{}{x}(\rho u^2+p)&=0\\
\pder{}{t}\left(\dfrac{1}{2}\rho u^2+\rho \epsilon\right)+\pder{}{x}\left[u\left(\dfrac{1}{2}\rho u^2+\rho \epsilon+p\right)\right]&=0.
\end{align}
\end{subequations}
These can be expressed as
\begin{align}
\label{eqn:euler-vec}
\pder{\vu}{t}+\pder{\vf(\vu)}{x}=0,
\end{align}
where $\vu$, $\vf$ are the conservative variables and their associated fluxes respectively:
\begin{align}
\vu=
\begin{pmatrix}
\rho\\
\rho u\\
E
\end{pmatrix}
\end{align}

\begin{align}\label{eqn:vflux}
\vf=
\begin{pmatrix}
\rho u\\
\rho u^2+p\\
u\rho H
\end{pmatrix}
=
\begin{pmatrix}
U_2\\
U_2^2/U_1+p(\vu)\\
U_2/U_1(U_3+p(\vu))
\end{pmatrix},
\end{align}
where $E$ is the total energy density, $H$ is the total specific enthalpy
\begin{align}
E&\equiv\dfrac{1}{2}\rho u^2+\rho\epsilon\\
H&\equiv \dfrac{1}{2}u^2+\epsilon+\dfrac{p}{\rho},
\end{align}
and $U_i$ are the components of $\vu$.
We can then linearize equation \ref{eqn:euler-vec} into
\begin{align}
\label{eqn:euler-vec2}
\pder{\vu}{t}+\va(\vu)\pder{\vu}{x}=0,
\end{align}
where $\va$ is the Jacobian of $\vf$
\begin{align}
\va(\vu)\equiv\pder{\vf}{\vu}=
\begin{pmatrix}
 0 & 1 & 0 \\
 p_1-\left(\frac{U_2}{U_1}\right)^2 & p_2+\frac{2 U_2}{U_1} & p_3 \\
 \frac{U_2 p_1}{U_1}-\frac{U_2 (p+U_3)}{U_1^2} & \frac{p+U_3}{U_1}+\frac{p_2 U_2}{U_1} & \frac{U_2 (p_3+1)}{U_1} \\
\end{pmatrix},
\end{align}
where 
\begin{align}
\label{eqn:p-der}
p_i\equiv\pder{p}{U_i}=p_\rho\pder{\rho}{U_i}+p_\epsilon\pder{\epsilon}{U_i},
\end{align}
and
\begin{align}
\rho=U_1,\,\,\epsilon=\dfrac{U_3}{U_1}-\dfrac{1}{2}\left(\dfrac{U_2}{U_1}\right)^2.
\end{align}

\textit{Side note}: Our present analysis should extend to the case of a more general EoS of the form $p(\rho,\epsilon,\bm{\psi})$ as long as
$p_\rho>0$, $p_\epsilon>0$,
and Eqn. \ref{eqn:p-der} holds, implying
\begin{align}
\pder{\bm{\psi}}{\vu}=\pder{\bm{\psi}}{\rho}\pder{\rho}{\vu}+\pder{\bm{\psi}}{\epsilon}\pder{\epsilon}{\vu},\,\text{ or }\,\pder{\bm{\psi}}{\vu}=0.
\end{align}
For example $\bm{\psi}$ could be some collection of passively advected scalars.

We can write $\va$ in a more manageable form by utilizing the relation below
\begin{align}
\label{eqn:asq}
a^2\equiv \pder{p}{\rho}{s}=\dfrac{p}{\rho^2}p_\epsilon+p_\rho=\dfrac{1}{\epsilon_p}\left(\dfrac{p}{\rho^2}-\epsilon_\rho\right),
\end{align}
to remove all instances of $p_\rho$. Where $a$ is the adiabatic sound speed.
\begin{align}
\label{eqn:A}
\va=
\begin{pmatrix}
 0 & 1 & 0 \\
 a^2+\frac{\left(u^2-H\right) p_{\epsilon }}{\rho
   }-u^2 & u \left(2-\frac{p_{\epsilon }}{\rho
   }\right) & \frac{p_{\epsilon }}{\rho } \\
 \frac{u \left(a^2 \rho -H \left(p_{\epsilon }+\rho
   \right)+u^2 p_{\epsilon }\right)}{\rho } &
   H-\frac{u^2 p_{\epsilon }}{\rho } & \frac{u
   \left(p_{\epsilon }+\rho \right)}{\rho } \\
\end{pmatrix}.
\end{align} 
The eigenvalues along with the associated left and right eigenvectors of $\va$ are,
\begin{subequations}
\begin{align}
\label{eqn:eigensys}
\lambda_1&=u-a
&\mathbf{l}^{(1)}&=\left(\dfrac{p_\epsilon}{\rho}(H-u^2)-a(a+u),\;u\dfrac{p_\epsilon}{\rho}+a,\;-\dfrac{p_\epsilon}{\rho}\right)
&\mathbf{r}^{(1)}&= 
\begin{pmatrix}
 1 \\
 u-a \\
 H-a u
\end{pmatrix}\\
\lambda_2&=u
&\mathbf{l}^{(2)}&=\left(H-u^2,\;u,\;-1\right)
&\mathbf{r}^{(2)}&=
\begin{pmatrix}
 1 \\
 u \\
 H-\frac{a^2 \rho }{p_{\epsilon }}
\end{pmatrix}\\
\lambda_3&=u+a
&\mathbf{l}^{(3)}&=\left(\dfrac{p_\epsilon}{\rho}(H-u^2)-a(a-u),\;u\dfrac{p_\epsilon}{\rho}-a,\;-\dfrac{p_\epsilon}{\rho}\right)
&\mathbf{r}^{(3)}&=
\begin{pmatrix}
 1 \\
 u+a \\
 H+a u
\end{pmatrix}.
\end{align}
\end{subequations}
Surprisingly, the only difference in the right eigenvectors $\left(\mathbf{r^{(i)}}\right)$ compared to the case of a perfect gas EoS (i.e. $\gamma$-law) is $\mathbf{r}_3^{(2)}$, in which $H-a^2\rho/p_e$ simplifies to $u^2/2$.

\subsection{Requirements on the EoS to keep the Euler equations hyperbolic}\label{sec:hyperbolicity}
It is crucial to make sure that the linearized Euler equations are hyperbolic with the EoS since Riemann solvers are based on Riemann invariants and Riemann invariants exist only when the PDEs are hyperbolic. In section \ref{sec:conserveform}, we have listed the eigenvalues of the Jacobian of the quasilinear system. The quasilinear system is hyperbolic if and only if all the eigenvalues are real and the Jacobian is diagonalizable. It is apparent that when $a^{2}>0$, the three eigenvalues are distinct and the quasilinear system is hyperbolic.

In equation \ref{eqn:asq}, $\epsilon_{p}>0$ because the specific internal energy is usually positively related to the pressure when the density is held constant. Therefore, the requirements of the EoS to keep the Euler equations hyperbolic now reduce to
\begin{equation}
\frac{p}{\rho^{2}}-\epsilon_{\rho}>0
\end{equation}
In this research, the EoS that we use has $\epsilon_{\rho}<0$, thus the linearized Euler equations are hyperbolic.

\subsection{Jump condition for the contact discontinuity}
For quasilinear hyperbolic systems, the Riemann invariants admitted by \ref{eqn:euler-vec2} follow the following ODEs \citep{jeffrey1976,toro2013}
\begin{align}
\dfrac{\dd U_i}{\mathbf{r}^{(k)}_i}=\dfrac{\dd U_j}{\mathbf{r}^{(k)}_j}\,\,\,\text{for all } i,j\in \left\lbrace 1,2,3 \right\rbrace.
\end{align}
The middle characteristic ($k=2$) is the contact discontinuity, and is the only characteristic for which we do not need to worry about non-linear shocks. From $i=1$ and $j=2$ we get
\begin{align}\label{eqn:jumpdu}
\dd u=0,
\end{align}
i.e. constant velocity across the contact discontinuity.
Using the above in combination with $i=1$ and $j=3$ we retrieve
\begin{align}
\left(H-\dfrac{a^2\rho}{p_\epsilon}\right) \dd\rho=\dd E=\rho\dd\epsilon+\left(\epsilon+\dfrac{1}{2}u^2\right)\dd\rho.
\end{align}
By recalling Eqn. \ref{eqn:asq} we can further simplify this to
\begin{align}\label{eqn:jumpdp}
p_\rho\dd\rho+p_\epsilon\dd\epsilon=\dd p=0,
\end{align}
i.e. pressure is constant across the contact discontinuity.

\section{Idealized Hydrogen EoS}\label{sec:eos}
To test our Riemann solvers, we consider a simple, but useful, EoS which contains a disassociation transition and an ionization transition. Our simple EoS describes a plasma with only four species:  molecular hydrogen (\ch{H2}), neutral hydrogen (\ch{H}) protons/ionized hydrogen (\ch{H+}), and electrons (\ch{e-}). Note that we do not consider the vibrational and rotational energy in \ch{H2} \cite{tomida2013,tomida2015} here because: (1) the partition function of \ch{H2} will be very complicated if we take these two energies into consideration and will kill a fast algorithm we will introduce hereafter, (2) these two kinds of energies are not significant compared to the binding energies, (3) the EoS is still monotone thus all the conclusions regarding the general EoS Riemann solvers hold, and (4) the ratio of orthohydrogen to parahydrogen may still be an assumption. However, we admit that it is desirable to include the vibrational and rotational energies, as this is necessary to recover the correct variation of $\Gamma$ below the molecular dissociation temperature. This added complexity is computationally prohibitive, requiring the interpolation of EoS. The EoS study here is crude in the sense that we only require it to manifest the monotonicity. We also want to present a fast algorithm in the case if we only consider the disassociation and the ionization binding energies. A more thorough study of the error of ignoring the vibrational and the rotational energies should be done in the future, but due to the page limit, we cannot present it here. We shall assume local thermal equilibrium (LTE), ideal gas law ($p=nkT$) and consider only two reactions in equilibrium
\begin{align}\label{eqn:H2}
\ch{H2} &\rightleftharpoons 2\ch{H},\\
\label{eqn:Hion}
\ch{H} &\rightleftharpoons \ch{H+}+\ch{e-}.
\end{align}

To derive this EoS we need to know the relevant partition functions. $Z_{i}$ is the partition function per volume for species $i$ (henceforth partition function will be used to mean partition function per volume). The partition function can be broken down into parts
\begin{equation}
Z_{i}= Z_{i}^{\rm bound} \times Z_{i}^{\rm nuc} \times Z_{i}^{\rm elec} \times Z_{i}^{\rm tr} \times Z_{i}^{\rm exct}.
\end{equation}
These parts are: internal bound states $Z_{i}^{\rm bound}$, nuclear spin $Z_{i}^{\rm nuc}$, electron spin and angular momentum $Z_{i}^{\rm elec}$, translation $Z_{i}^{\rm tr}$, and excitation $Z_{i}^{\rm exct}$. To further simplify the problem we shall assume that
\begin{align}
Z_{i}^{\rm bound}=1\;\;\text{for all }i.
\end{align}
While this is not technically true for \ch{H}, it is a relatively small effect on the EoS, 
but the lack of rotational and vibrational states for $\ch{H}_2$ effects $\Gamma$.
However, this EoS is meant as a simple proof of concept and not for high precision applications.

We now list the partition functions for all the species. The partition functions not explicitly described are unity. The translational partition function has the same form for all species
\begin{equation}
Z_{i}^{\rm tr}=\left(\dfrac{2\pi m_{i}kT}{h^2}\right)^{3/2},
\end{equation}
where $m_i$ is the mass of the $i$th species and $h$ is the Planck constant.

The remaining non-trivial partition functions are 
\begin{align}
Z_{\ch{H2}}^{\rm elec}&=2,\\
Z_{\ch{H}}^{\rm nuc}&=2,\\
Z_{\ch{H}}^{\rm elec}&=2,\\
Z_{\ch{H}}^{\rm exct}&=\exp\left(-\frac{\phi_{\rm dis}}{2kT}\right),\\
Z_{\ch{H+}}^{\rm nuc}&=2,\\
Z_{\ch{H}}^{\rm elec}&=2,\\
Z_{\ch{H+}}^{\rm exct}&=\exp\left(-\frac{\phi_{\rm dis}+2\phi_{\rm ion}}{2kT}\right),\\
Z_{\ch{e-}}^{\rm elec}&=2,
\end{align}
where $\phi_{\rm dis}=7.17\times10^{-12}$erg \citep{liu2009} and $\phi_{\rm ion}=2.18\times10^{-11}$erg while all the unspecified partition functions are unity.

\begin{figure}
    \centering
    \includegraphics[width=0.8\columnwidth]{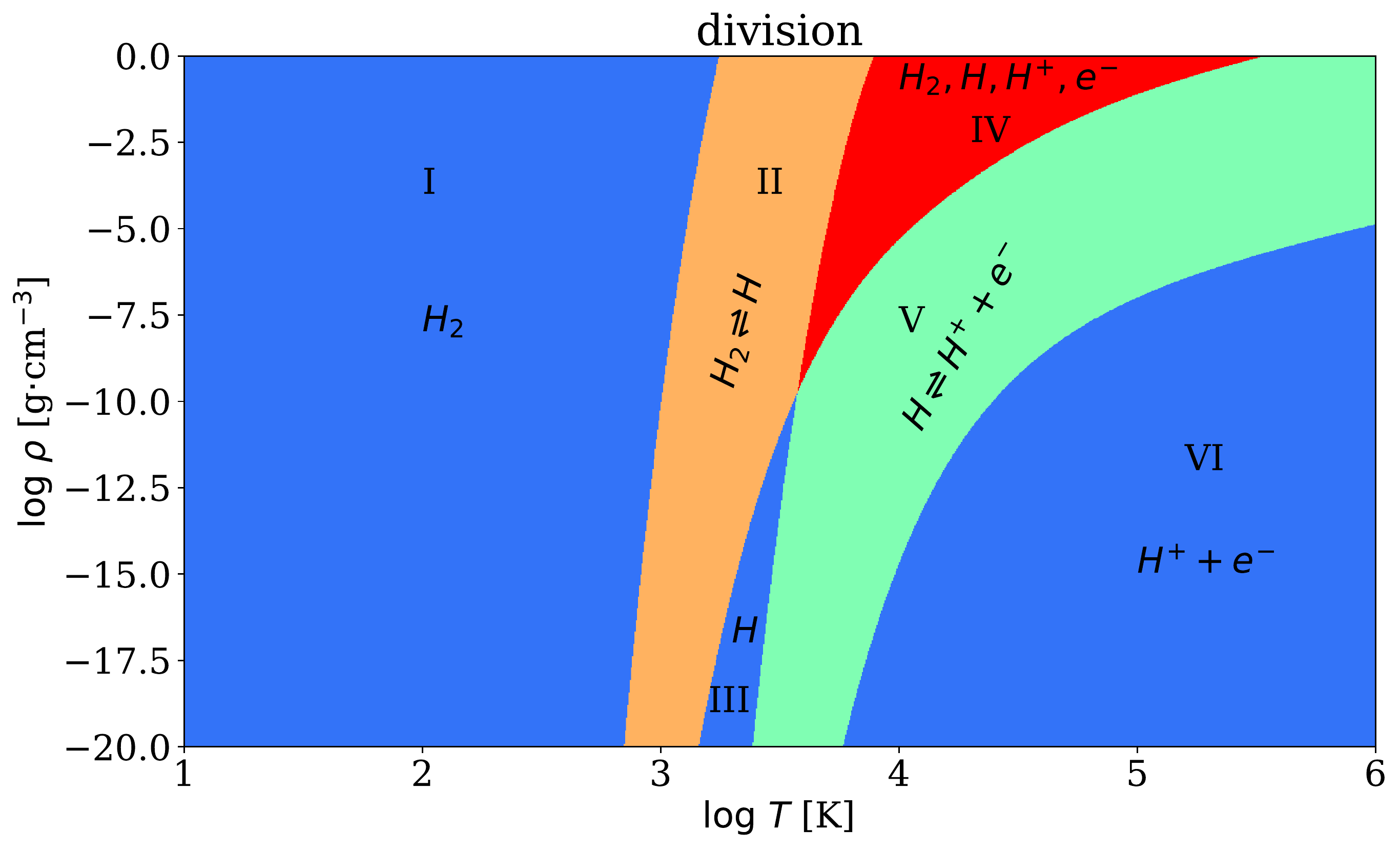}
    \caption{Division of the EoS in $(\rho,T)$.}
    \label{fig:division}
\end{figure}

To compute the EoS we must first solve number conservation, Saha equations corresponding to Equations \ref{eqn:H2} and \ref{eqn:Hion}, and charge neutrality to get the number density of all species:
\begin{align}
2n_{\ch{H2}}+n_{\ch{H}}+n_{\ch{H+}}&=n_{\rm Htot}=\frac{\rho}{m_h},\label{eqn:nconserv}\\ \frac{n_{\ch{H}}^2}{n_{\ch{H2}}}&=\frac{Z_{\ch{H}}^2}{Z_{\ch{H2}}}=q_{\rm dis},\label{eqn:qdis}\\
\frac{n_{\ch{H+}}n_{\ch{e-}}}{n_{\ch{H}}}&=\frac{Z_{\ch{H+}}Z_{\ch{e-}}}{Z_{\ch{H}}}=q_{\rm ion},\label{eqn:qion}\\
n_{\ch{H+}}-n_{\ch{e-}}&=0\label{eqn:econserv}
\end{align}
respectively, where $n_{i}$ is the number density of the $i$th species. Substitute Equation \ref{eqn:qdis}, \ref{eqn:qion}, and \ref{eqn:econserv} into \ref{eqn:nconserv}, we can obtain a quartic equation of $n_{\ch{H+}}$,
\begin{equation}\label{eqn:quartic}
    \frac{2n_{\ch{H+}}^4}{q_{\rm ion}q_{\rm dis}^2}+\frac{n_{\ch{H+}}^2}{q_{\rm ion}}+n_{\ch{H+}}=n_{\rm Htot}.
\end{equation}
Once $n_{\ch{H+}}$ is determined, the number density of other species can be obtained by solving Equation \ref{eqn:qdis}, \ref{eqn:qion}, and \ref{eqn:econserv}. Fourth degree polynomials are the highest degree where an explicit solution is guarantied. However, solving Equation \ref{eqn:quartic} in computer with finite accuracy will not always give satisfactory result when $q_{\rm dis}$ and $q_{\rm ion}$ are too large or too small. Besides, solving a quartic equation is much more expensive than solving a quadratic equation. To improve the computational efficiency, we need to minimize the solution of the quartic equation. Therefore, we divide the EoS in density and temperature space $(\rho,T)$ into six regions with the following division criterion: if the concentration of \ch{H} (Equation \ref{eqn:concentration1}-\ref{eqn:concentration3}) in any species is less than $10^{-6}$, we assume such species does not present. The division result is shown in Figure \ref{fig:division}.

In the meantime, we get four curves in $(\rho,T)$ that delineate these six regions. They are:
\begin{itemize}
    \item $T_{1}(\rho)$ separates the \ch{H2} only region and the multi-species region.
    \item $T_{2}(\rho)$ separates the \ch{H2} present and non-present region.
    \item $T_{3}(\rho)$ separates the \ch{H+} present and non-present region.
    \item $T_{4}(\rho)$ separates the \ch{H+} and \ch{e-} only region and the multi-species region.
\end{itemize}
Note that $T_{2}(\rho)<T_{3}(\rho)$ in region \romcap{3} but $T_{2}(\rho)>T_{3}(\rho)$ in region \romcap{4}. These four $(\rho,T)$ curves will be used in calculating the temperature with $(\rho,p)$ and $(\rho,\epsilon)$.

The blue color indicates the regions with single species. When $(\rho,T)$ falls into these regions, we can easily calculate the number density of that species.

If $(\rho,T)$ belongs to \romcap{2}, we assume $n_{\ch{H+}}=n_{\ch{e-}}=0$. We solve Equation \ref{eqn:nconserv} and \ref{eqn:qdis}. Substitute Equation \ref{eqn:qdis} into \ref{eqn:nconserv}, we can get a quadratic equation of $n_{\ch{H}}$,
\begin{equation}\label{eqn:IIa}
    \frac{2n_{\ch{H}}^2}{q_{\rm dis}}+n_{\ch{H}}=n_{\rm Htot}
\end{equation}
and the physical solution of Equation \ref{eqn:IIa} is,
\begin{equation}
    n_{\ch{H}}=\frac{2q_{\rm dis}n_{\rm Htot}}{q_{\rm dis}+\sqrt{q_{\rm dis}^{2}+8q_{\rm dis}n_{\rm Htot}}}.
\end{equation}

If $(\rho,T)$ belongs to \romcap{4}, we assume all four species present and we have to solve Equation \ref{eqn:quartic} and then Equation \ref{eqn:qdis}, \ref{eqn:qion}, and \ref{eqn:econserv}.

If $(\rho,T)$ belongs to \romcap{5}, we assume $n_{\ch{H2}}=0$. We solve Equation \ref{eqn:nconserv}, \ref{eqn:qion}, and \ref{eqn:econserv}. Substitute Equation \ref{eqn:qion} and \ref{eqn:econserv} into \ref{eqn:nconserv}, we can get a quadratic equation of $n_{\ch{H+}}$,
\begin{equation}\label{eqn:IIc}
    \frac{n_{\ch{H+}}^2}{q_{\rm ion}}+n_{\ch{H+}}=n_{\rm Htot}
\end{equation}
and the physical solution of Equation \ref{eqn:IIc} is,
\begin{equation}
    n_{\ch{H}}=\frac{2q_{\rm ion}n_{\rm Htot}}{q_{\rm ion}+\sqrt{q_{\rm ion}^{2}+4q_{\rm ion}n_{\rm Htot}}}.
\end{equation}

After solving $n_{i}$, the pressure and specific internal energy can be calculated by,
\begin{equation}\label{eqn:eosp}
    p=\sum_{i}n_{i}kT
\end{equation}
\begin{equation}\label{eqn:eosepsilon}
    \epsilon=\frac{1}{\rho}\left(\sum_{i}n_{i}\xi_{i}\right)
\end{equation}
where
\begin{align}
    \xi_{\ch{H2}}&=3kT/2\\
    \xi_{\ch{H}}&=3kT/2+\phi_{\rm dis}/2\\
    \xi_{\ch{H+}}&=3kT/2+\phi_{\rm dis}/2+\phi_{\rm ion}\\
    \xi_{\ch{e-}}&=3kT/2
\end{align}
are the internal energy per particle of each species.

We also need to calculate the adiabatic sound speed, which is needed in the HLLC general EoS Riemann solver. If the specific entropy $s$ is differentiable, the adiabatic sound speed $a$ can be calculated by,
\begin{align}
    a=\sqrt{\left(\frac{\partial p}{\partial\rho}\right)_{s}}&=\sqrt{p_{\rho}-\frac{p_{T}s_{\rho}}{s_{T}}},
\end{align}
where
\begin{align}
    s&=\sum_{i}\frac{kn_{i}}{\rho}\left(1+\frac{d\ln Z_{i}}{d\ln T}-\ln\frac{n_{i}}{Z_{i}}\right)\\
    s_T&=\left(\dfrac{\partial s}{\partial T}\right)_T\\
    s_\rho&=\left(\dfrac{\partial s}{\partial \rho}\right)_\rho\\
    p_T&=\left(\dfrac{\partial p}{\partial T}\right)_T\\
    p_\rho&=\left(\dfrac{\partial p}{\partial \rho}\right)_\rho.
\end{align}
Although there are many derivatives in the expression of $S_{T},S_{\rho},p_{T}$, and $p_{\rho}$, the derivatives of the partition functions are simply,
\begin{align}
    d\ln Z_{\ch{H2}}/d\ln T&=3/2\\
    d\ln Z_{\ch{H}}/d\ln T&=3/2+\phi_{\rm dis}/(2kT)\\
    d\ln Z_{\ch{H+}}/d\ln T&=3/2+(\phi_{\rm dis}+2\phi_{\rm ion})/(2kT)\\
    d\ln Z_{\ch{e-}}/d\ln T&=3/2\\
    d^{2}\ln Z_{\ch{H2}}/d\ln T^{2}&=0\\
    d^{2}\ln Z_{\ch{H}}/d\ln T^{2}&=-\phi_{\rm dis}/(2kT)\\
    d^{2}\ln Z_{\ch{H+}}/d\ln T^{2}&=-(\phi_{\rm dis}+2\phi_{\rm ion})/(2kT)\\
    d^{2}\ln Z_{\ch{e-}}/d\ln T^{2}&=0
\end{align}
This would not be the case if one consider the vibrational and rotational degree of freedom of \ch{H2} as in \cite{tomida2013,tomida2015}.

The only derivatives that we need to be calculated are $\left(\frac{\partial\ln n_{i}}{\partial\ln T}\right)_{\rho}$ and $\left(\frac{\partial\ln n_{i}}{\partial\ln\rho}\right)_{T}$. When all species present, i.e., $(\rho,T)\in$\romcap{4}, let us re-write Equation \ref{eqn:nconserv}-\ref{eqn:qion} in log form while using \ref{eqn:econserv},
\begin{align}\label{eqn:dlnndlnx}
\begin{pmatrix}
2n_{\ch{H2}}  &  n_{\ch{H}}  &  n_{\ch{n+}}  \\
-1  &  2  &  0  \\
0  &  -1  &  2
\end{pmatrix}
\begin{pmatrix}
d\ln n_{\ch{H2}}\\
d\ln n_{\ch{H}}\\
d\ln n_{\ch{H+}}
\end{pmatrix}=
\begin{pmatrix}
n_{\rm Htot}d\ln n_{\rm Htot}\\
d\ln q_{\rm dis}\\
d\ln q_{\rm ion}
\end{pmatrix}
\end{align}
Let the right-hand side of Equation \ref{eqn:dlnndlnx} be $\mathbf{L}$. Then,
\begin{align}
\left(\frac{\partial\mathbf{L}}{\partial\ln T}\right)_{\rho}=
\begin{pmatrix}
L_{T1}\\
L_{T2}\\
L_{T3}
\end{pmatrix}=
\begin{pmatrix}
0\\
2\xi_{\ch{H}}-\xi_{\ch{H2}}\\
\xi_{\ch{H+}}+\xi_{\ch{e-}}-\xi_{\ch{H}}
\end{pmatrix}
\end{align}
and,
\begin{align}
\left(\frac{\partial\mathbf{L}}{\partial\ln\rho}\right)_{T}=
\begin{pmatrix}
L_{\rho1}\\
L_{\rho2}\\
L_{\rho3}
\end{pmatrix}=
\begin{pmatrix}
n_{\rm Htot}\\
0\\
0
\end{pmatrix}
\end{align}

Then,
\begin{align}
\left(\frac{\partial\ln n_{\ch{H2}}}{\partial\ln T}\right)_{\rho}&=2\left(\frac{\partial\ln n_{\ch{H}}}{\partial\ln T}\right)_{\rho}-L_{T2}\\
\left(\frac{\partial\ln n_{\ch{H}}}{\partial\ln T}\right)_{\rho}&=2\left(\frac{\partial\ln n_{\ch{H+}}}{\partial\ln T}\right)_{\rho}-L_{T3}\\
\left(\frac{\partial\ln n_{\ch{H+}}}{\partial\ln T}\right)_{\rho}&=\frac{2n_{\ch{H2}}L_{T2}+(4n_{\ch{H2}}+n_{\ch{H}})L_{T3}}{8n_{\ch{H2}}+2n_{\ch{H}}+n_{\ch{H+}}}\\
\left(\frac{\partial\ln n_{\ch{e-}}}{\partial\ln T}\right)_{\rho}&=\left(\frac{\partial\ln n_{\ch{H+}}}{\partial\ln T}\right)_{\rho}
\end{align}
and
\begin{align}
\left(\frac{\partial\ln n_{\ch{H2}}}{\partial\ln\rho}\right)_{T}&=2\left(\frac{\partial\ln n_{\ch{H}}}{\partial\ln\rho}\right)_{T}\\
\left(\frac{\partial\ln n_{\ch{H}}}{\partial\ln\rho}\right)_{T}&=2\left(\frac{\partial\ln n_{\ch{H+}}}{\partial\ln\rho}\right)_{T}\\
\left(\frac{\partial\ln n_{\ch{H+}}}{\partial\ln\rho}\right)_{T}&=\frac{n_{\rm Htot}}{8n_{\ch{H2}}+2n_{\ch{H}}+n_{\ch{H+}}}\\
\left(\frac{\partial\ln n_{\ch{e-}}}{\partial\ln\rho}\right)_{T}&=\left(\frac{\partial\ln n_{\ch{H+}}}{\partial\ln\rho}\right)_{T}
\end{align}
When $(\rho,T)\in$\romcap{1}
\begin{equation}
    \left(\frac{\partial\ln n_{\ch{H2}}}{\partial\ln\rho}\right)_{T}=\frac{\rho}{n_{\ch{H2}}m_{\ch{H2}}}
\end{equation}
and all the other derivatives are set to be $0$. When $(\rho,T)\in$\romcap{3},
\begin{equation}
    \left(\frac{\partial\ln n_{\ch{H}}}{\partial\ln\rho}\right)_{T}=\frac{\rho}{n_{\ch{H}}m_{\ch{H}}}
\end{equation}
and all the other derivatives are set to be $0$. When $(\rho,T)\in$\romcap{5},
\begin{align}
\left(\frac{\partial\ln n_{\ch{H+}}}{\partial\ln\rho}\right)_{T}&=\frac{\rho}{n_{\ch{H+}}m_{\ch{H+}}}\\
\left(\frac{\partial\ln n_{\ch{e-}}}{\partial\ln\rho}\right)_{T}&=\left(\frac{\partial\ln n_{\ch{H+}}}{\partial\ln\rho}\right)_{T}
\end{align}
and all the other derivatives are set to be $0$. When $(\rho,T)\in$\romcap{2} or \romcap{5}, the derivatives of non-presenting species are set to be $0$ while the derivatives of existing species can be calculated by solving a rank 2 linear system following the same step as four species.

It is also beneficial and necessary to be able to calculate temperature for a given $(\rho,p)$ or $(\rho,\epsilon)$. We design a fast algorithm to find such temperature with the EoS that satisfy:
\begin{align}
    \left(\frac{\partial p}{\partial\rho}\right)_{T}&=p_{\rho}>0,  \label{eqn:monoton1}\\
    \left(\frac{\partial p}{\partial T}\right)_{\rho}&=p_{T}>0,  \label{eqn:monoton2}\\
    \left(\frac{\partial\epsilon}{\partial\rho}\right)_{T}&=\epsilon_{\rho}\le0,  \label{eqn:monoton3}\\
    \left(\frac{\partial\epsilon}{\partial T}\right)_{\rho}&=\epsilon_{T}>0.  \label{eqn:monoton4}
\end{align}
In fact, almost all ideal gas EoS satisfy these monotone conditions. We now prove that our four monotone conditions are sufficient to the Equation \ref{eqn:unique1} and \ref{eqn:unique2}.
\begin{equation}
    \left(\frac{\partial p}{\partial\rho}\right)_{\epsilon}=\left(\frac{\partial p}{\partial\rho}\right)_{T}+\left(\frac{\partial p}{\partial T}\right)_{\rho}\left(\frac{\partial T}{\partial\rho}\right)_{\epsilon}>0
\end{equation}
using Equation \ref{eqn:monoton1}, \ref{eqn:monoton2}, and
\begin{equation}\label{sec:trhoepsilon}
    \left(\frac{\partial T}{\partial\rho}\right)_{\epsilon}=-\left(\frac{\partial T}{\partial\epsilon}\right)_{\rho}\bigg/\left(\frac{\partial\rho}{\partial\epsilon}\right)_{T}\ge0.
\end{equation}
Note that when the inequality \ref{eqn:monoton3} becomes an equality, Equation \ref{sec:trhoepsilon} become undefined. This is because $\epsilon(\rho,T)$ is reduced to $\epsilon(T)$. For simplicity, we define this situation to be $0$. The EoS of perfect gas is a such example. On the other hand,
\begin{equation}
    \left(\frac{\partial p}{\partial\epsilon}\right)_{\rho}=\left(\frac{\partial p}{\partial T}\right)_{\rho}\left(\frac{\partial T}{\partial\epsilon}\right)_{\rho}>0
\end{equation}
because of Equation \ref{eqn:monoton2} and \ref{eqn:monoton4}.

The steps for solving $T=T(\rho,p)$ are,
\begin{enumerate}[Step 1]
    \item With the given $\rho$, find the four temperature $T_{1}(\rho)$, $T_{2}(\rho)$, $T_{3}(\rho)$, and $T_{4}(\rho)$ by linear interpolation. Calculate the corresponding $p_{i}$ by solving the species number and ideal gas law $p_{i}=n(T_{i})kT_{i}$.
    \item If $p\le p_{1}$, the gas is in purely molecular state and
    \begin{equation}
        T=\frac{p}{n_{\ch{H2}}k}=\frac{pm_{\ch{H2}}}{\rho k}.
    \end{equation}
    If $p_{2}\le p\le p_{3}$, the gas is in fully atomic state and
    \begin{equation}
        T=\frac{p}{n_{\ch{H}}k}=\frac{pm_{\ch{H}}}{\rho k}.
    \end{equation}
    If $p\ge p_{4}$, the gas is in fully ionized state and
    \begin{equation}
        T=\frac{p}{2n_{\ch{H+}}k}=\frac{pm_{\ch{H+}}}{2\rho k}.
    \end{equation}
    \item Otherwise, find the lower ($p_{\rm l}$) and upper ($p_{\rm u}$) limit of the pressure of the region that $(\rho,p)$ belongs to. $p_{\rm l}<p<p_{\rm u}$, use bisection method to determine $T\in[T_{l},T_{u}]$ and dynamically update the lower and upper bound ($T_{\rm l}$ and $T_{\rm u}$). In this research, we define the solution as $T=(T_{\rm l}+T_{\rm u})/2$ when $2(T_{\rm u}-T_{\rm l})/(T_{\rm u}+T_{\rm l})<10^{-5}$.
\end{enumerate}
The steps for solving $T=T(\rho,\epsilon)$ follow the same logic as solving $T=T(\rho,p)$.

\begin{figure}
	\centering
    \includegraphics[width=0.8\textwidth]{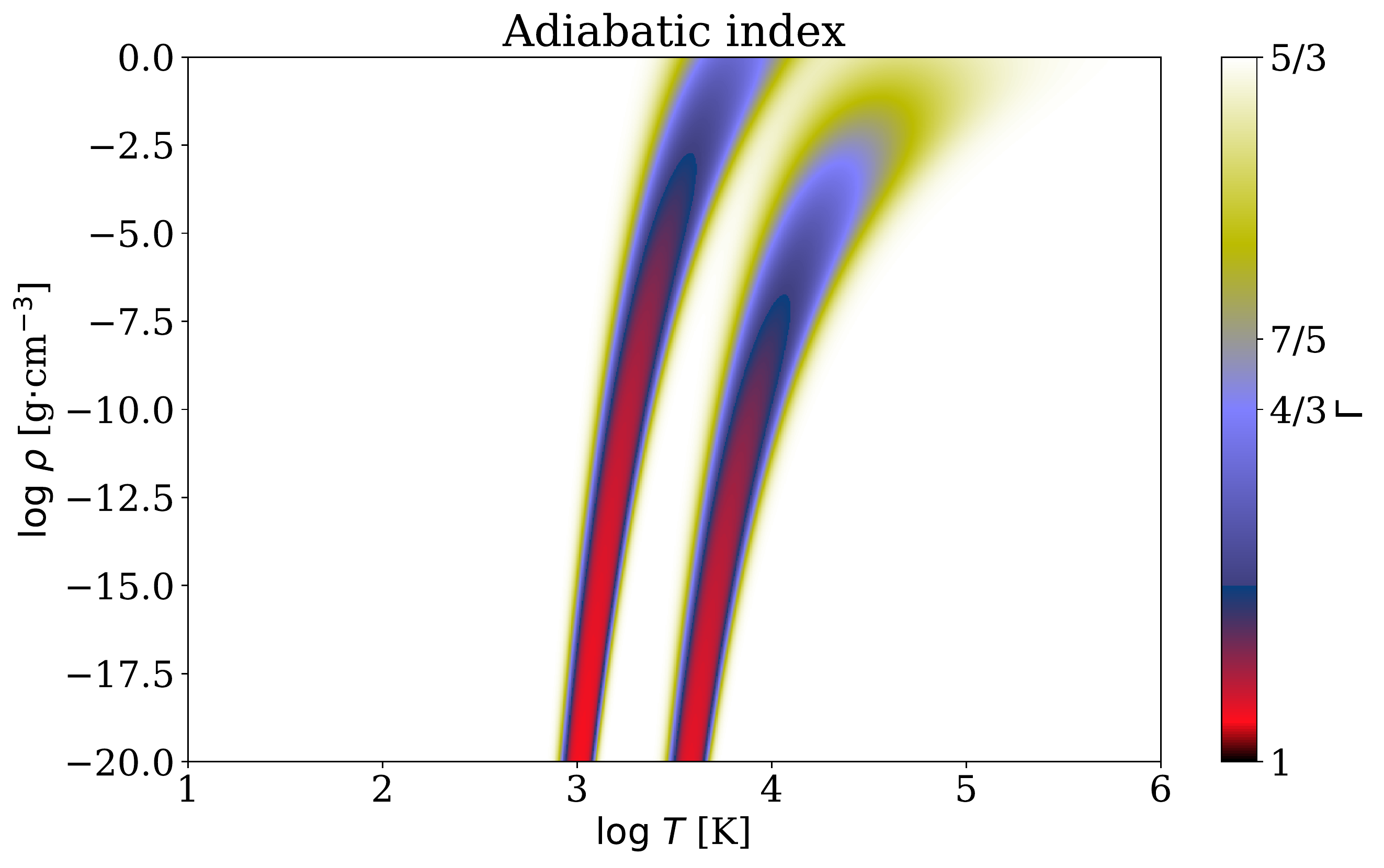}
    \caption{The table of the adiabatic index $\Gamma$. The $x$ axis is the temperature in $\log_{10}$ scale and the $y$ axis is the density in $\log_{10}$ scale. The two colored bands are where the disassociation of molecular hydrogen and the ionization of atomic hydrogen happen.}
    \label{fig:gamma}
\end{figure}

\begin{figure}
    \centering
    \includegraphics[width=0.8\textwidth]{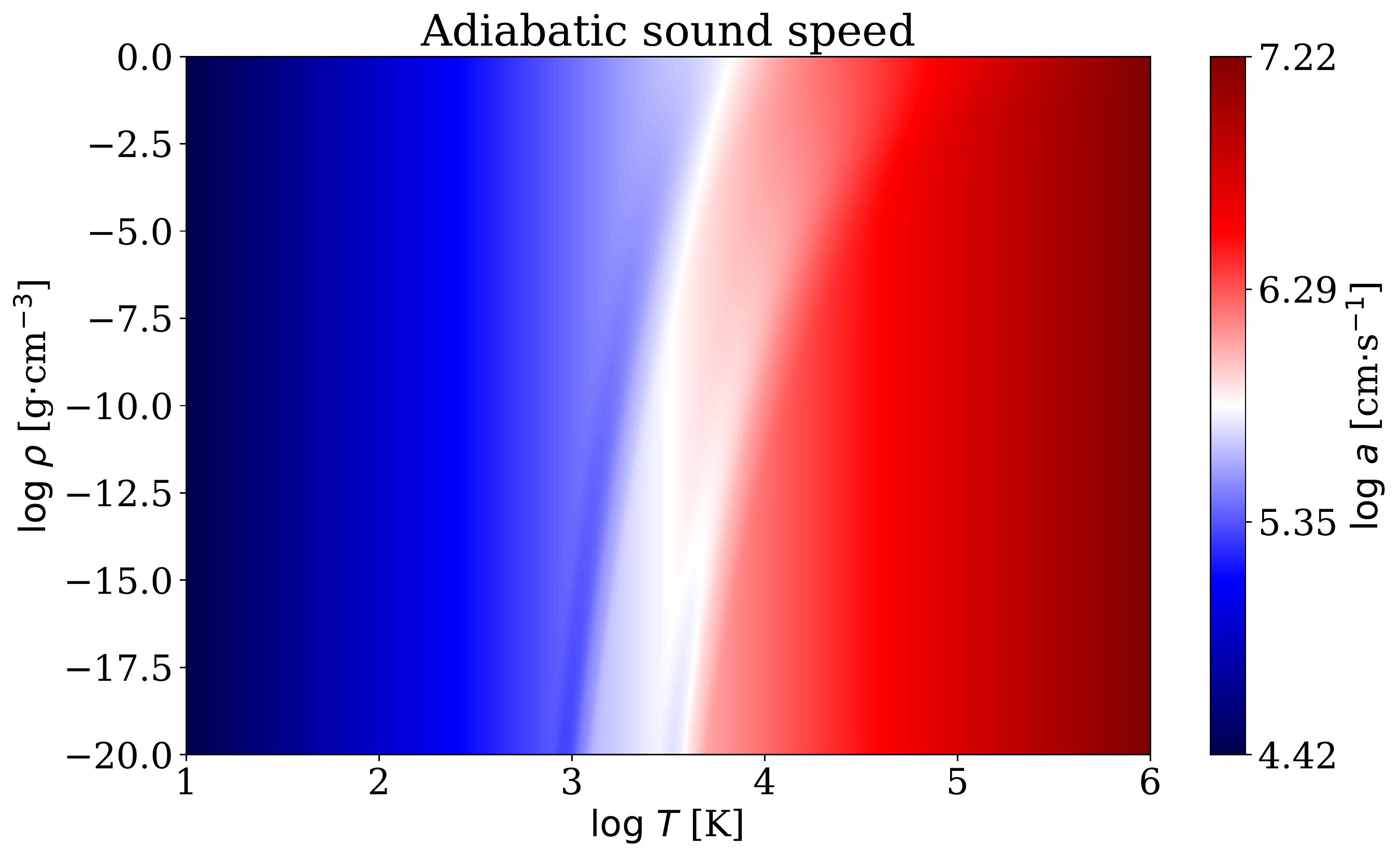}
    \caption{$\log a$ in $\log\rho$ and $\log T$. 
    }
    \label{fig:cs}
\end{figure}

\begin{figure}
	\centering
    \includegraphics[width=0.8\textwidth]{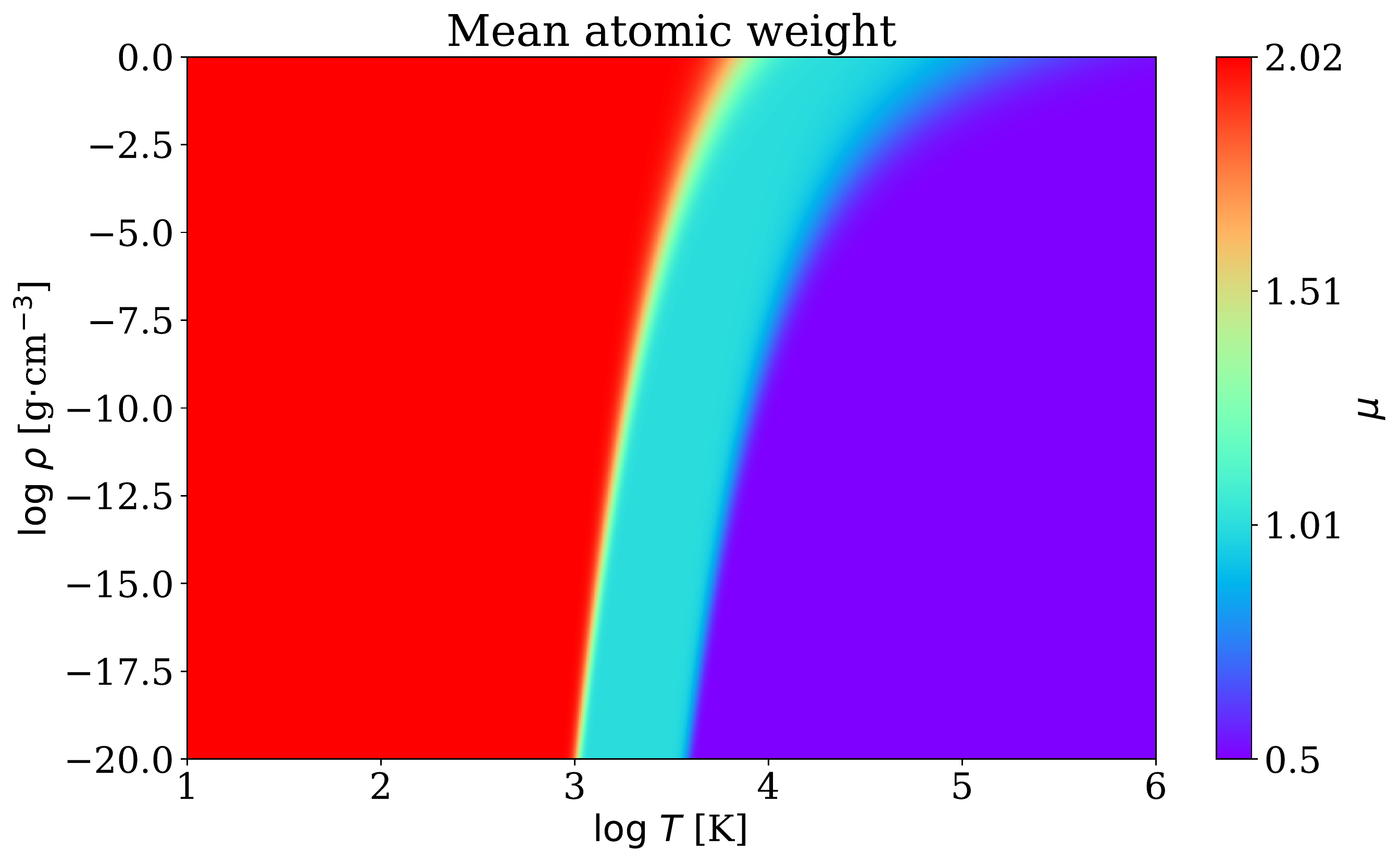}
    \caption{The table of the mean atomic weight $\mu$.}
    \label{fig:mu}
\end{figure}

\begin{figure}
	\centering
    \includegraphics[width=0.8\textwidth]{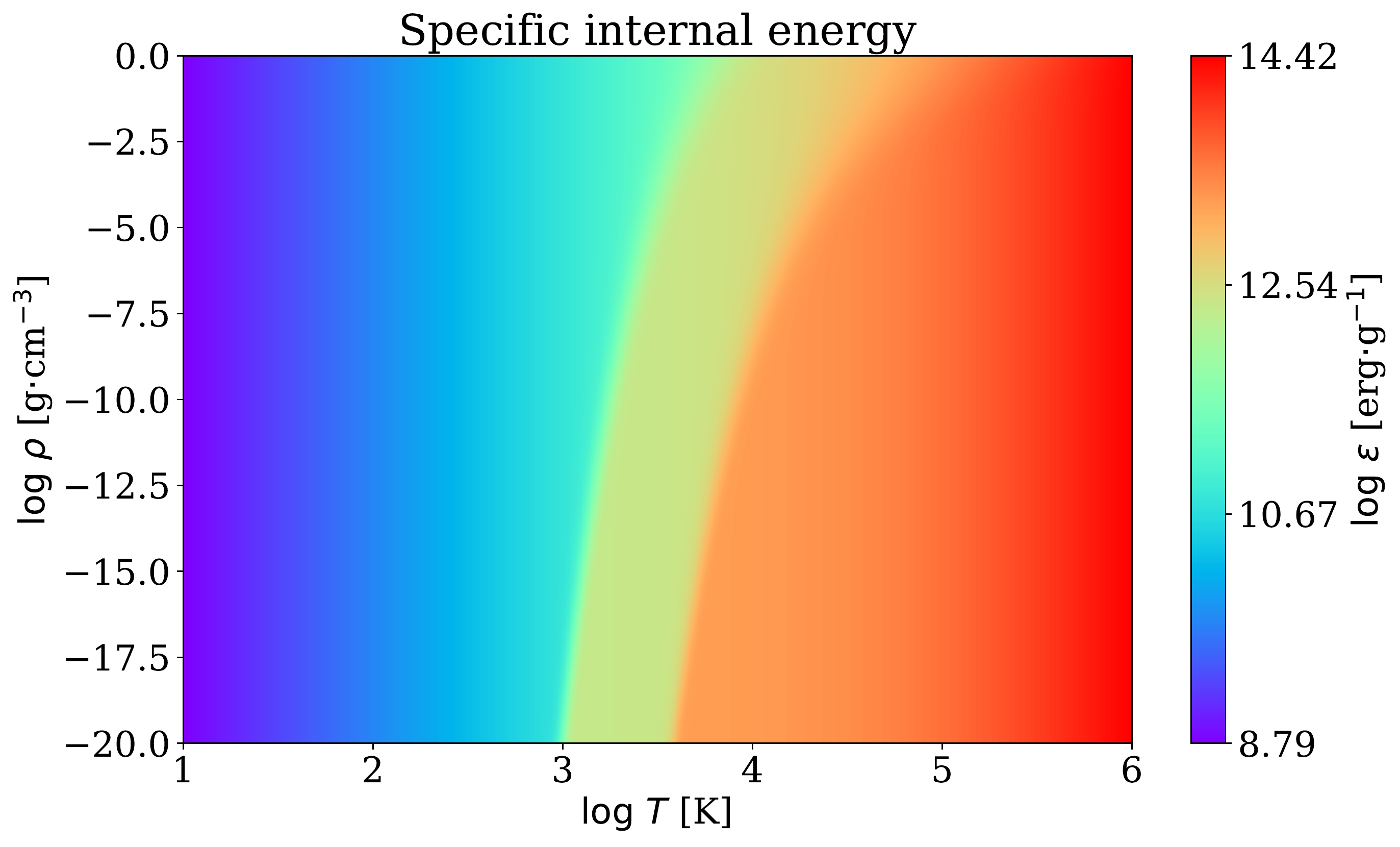}
    \caption{The table of the specific internal energy $\epsilon$ in the unit of erg$\cdot$g$^{-1}$ in $log_{10}$ scale.}
    \label{fig:eg}
\end{figure}

\begin{figure}
	\centering
    \includegraphics[width=0.8\textwidth]{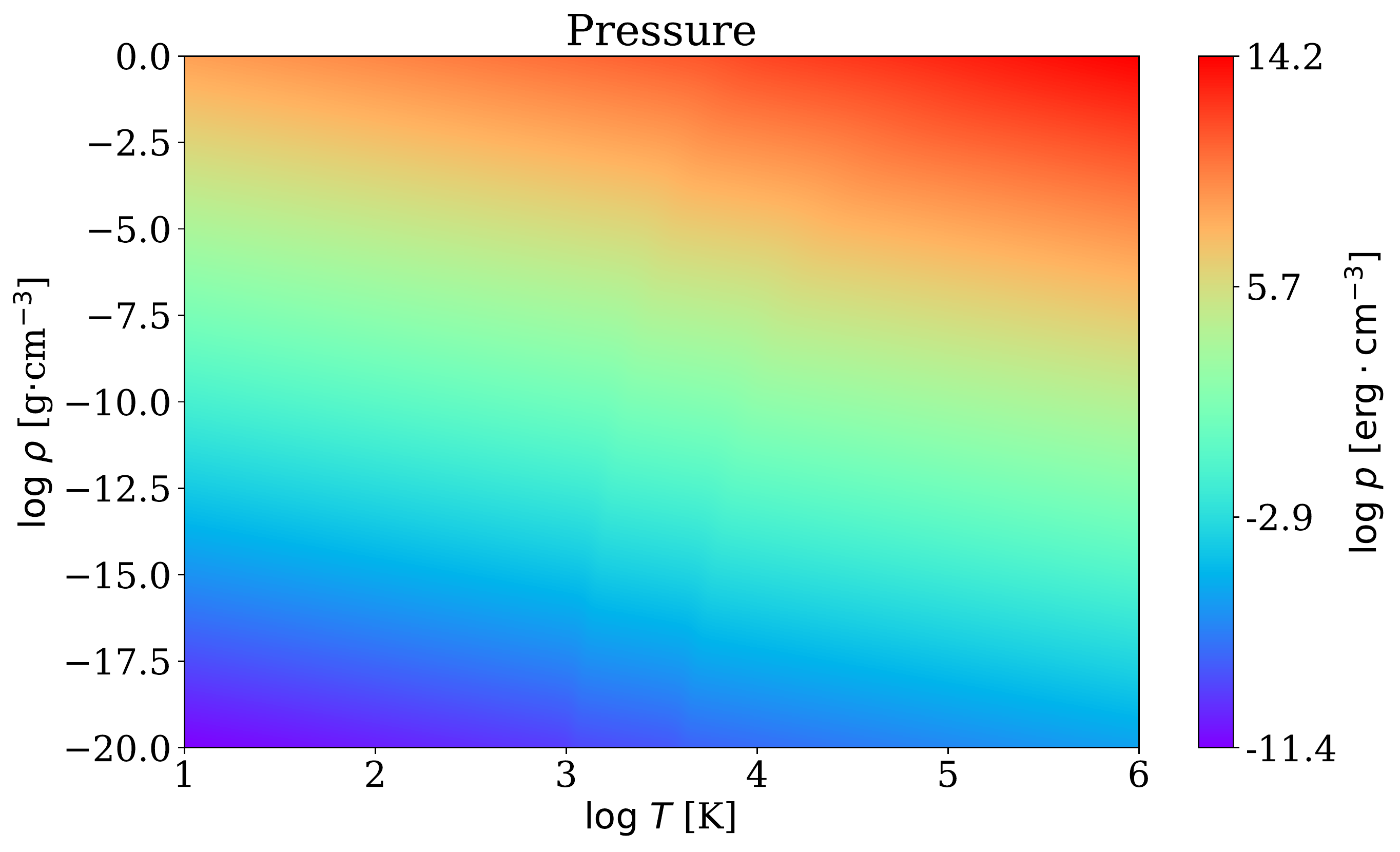}
    \caption{The table of the pressure $p$ in the unit of bar in $log_{10}$ scale.}
    \label{fig:p}
\end{figure}

We present a set of example EoS tables in Figure \ref{fig:gamma}, \ref{fig:cs}, \ref{fig:mu}, \ref{fig:eg}, and \ref{fig:p}. We show $a,\ \Gamma,\ \mu,\ \epsilon$, and $p$ in temperature range $T\in[10^{1},10^{6}]$ with $d\log T=0.005$ and $\rho\in[10^{-20},10^{0}]$ with $d\log\rho=0.02$, where $T$ and $\rho$ are given in units of K and \gcmc, respectively.

\end{appendices}

\section*{References}

\bibliography{main}

\end{document}